\RequirePackage{snapshot}
\documentclass[prb,aps,reprint,groupedaddress,showpacs]{revtex4-1}

\usepackage{natbib}
\usepackage[dvips]{graphicx}
\usepackage{amsmath}
\usepackage{amsfonts}
\usepackage{subfigure}
\usepackage{psfrag}
\usepackage{color} 
\newcommand{\al}{\alpha}
\newcommand{\eps}{\epsilon}
\newcommand{\be}{\begin{equation}}
\newcommand{\ee}{\end{equation}}
\newcommand{\bea}{\begin{eqnarray}}
\newcommand{\eea}{\end{eqnarray}}

\begin{document} 

\title{Adiabatic quantum pumping through surface
  states in 3D topological insulators }

\author{M. Alos-Palop, Rakesh P. Tiwari and M. Blaauboer}

\affiliation{Delft University of Technology, Kavli Institute of
  Nanoscience, Department of Quantum Nanoscience, Lorentzweg 1, 2628
  CJ Delft, The Netherlands.}

\date{\today}

\begin{abstract}
We investigate adiabatic quantum pumping of Dirac fermions on the
surface of a strong 3D topological insulator. Two different geometries
are studied in detail, a normal metal -- ferromagnetic -- normal metal
(NFN) junction and a ferromagnetic -- normal metal -- ferromagnetic
(FNF) junction. Using a scattering matrix approach, we first calculate
the tunneling conductance and then the adiabatically pumped current
using different pumping mechanisms for both types of junctions. We
explain the oscillatory behavior of the conductance by studying the
condition for resonant transmission in the junctions and find that
each time a new resonant mode appears in the transport window, the
pumped current diverges.
We also 
predict an experimentally distinguishable difference between the
pumped current and the rectified current.
\end{abstract}


\maketitle
\section{Introduction}

Recently, surface states in topological insulators have attracted a
lot of attention in the condensed-matter
community~\cite{Hasan2010}. Both in two-dimensional (e.g., HgTe) and
in three-dimensional (e.g., Bi$_2$Se$_3$) compounds with strong
spin-orbit interaction the topological phase has been demonstrated
experimentally~\cite{Konig2007,Hsieh2008,Xia2009a,Xia2009b}.  Although
these compounds are insulating in the bulk (since they have an energy
gap between the conduction band and the valance band), their surface
states support topological gapless excitations. In the simplest case
these low-energy excitations of a strong three-dimensional topological
insulator can be described by a single Dirac cone at the center of the
two-dimensional Brillouin zone ($\Gamma$
point)~\cite{Konig2007,Xia2009a,Hsieh2009a,Roushan2009,Hsieh2009b,*HZhang2009}.
The corresponding Hamiltonian is given by~\cite{Fu2008}
\begin{equation}
  \mathcal{H}_0=
  \hbar v_{F} \vec{\sigma} \cdot \vec{k} - \mu I.
  \label{eq:hamiltonian0}
\end{equation}
Here $\vec{\sigma}$ represents a vector whose three components are the
three Pauli spin matrices, $I$ represents a $2\times 2$ identity
matrix in spin space, $v_{F}$ is the Fermi velocity, and $\mu$ is the
chemical potential. The low-energy excitations of $\mathcal{H}_0$
[Eq.~(\ref{eq:hamiltonian0})] are topologically protected against
perturbations~\cite{TZhang2009}. This has prompted recent research on
the transport properties of surface Dirac fermions. For example, the
conductance and magnetotransport of Dirac fermions have been studied
in normal metal -- ferromagnet (NF), normal metal -- ferromagnetic --
normal metal (NFN) and arrays of NF junctions on the surface of a
topological insulator~\cite{Mondal2010a,Mondal2010b,Zhang2010},
suggesting the possibility of an engineered magnetic switch. An
anomalous magnetoresistance effect has been predicted in
ferromagnetic-ferromagnetic junctions~\cite{Yokoyama2010}. Also,
electron tunneling and magnetoresistance have been studied in
ferromagnetic -- normal metal -- ferromagnetic (FNF)
junctions~\cite{Wu2010,Salehi2011}, for which it has been predicted
that the conductance can be larger in the anti-parallel configuration
of the magnetizations of the two ferromagnetic regions than in the
parallel configuration. In addition, a large research effort has been
devoted to studying models which predict the existence of Majorana
fermion edge states at the interface between superconductors and
ferromagnets deposited on a topological
insulator~\cite{Fu2008,Akhmerov2009,Tanaka2009}.

In this article we investigate adiabatic quantum pumping of Dirac
fermions through edge states on the surface of a strong
three-dimensional topological insulator.  Quantum pumping refers to a
transport mechanism in meso- and nanoscale devices by which a finite
dc current is generated in the absence of an applied bias by periodic
modulations of at least two system parameters (typically gate voltages
or magnetic fields)~\cite{Buttiker1994,Brouwer1998,Spivak1995}. In
order for electrical transport to be adiabatic, the period of the
oscillatory driving signals has to be much longer than the dwell time
$\tau_{\rm dwell}$ of the electrons in the system, $ T = 2 \pi
\omega^{-1} \gg \tau_{\rm dwell} $. In the last decade, many different
aspects of quantum pumping have been theoretically investigated in a
diverse range of nanodevices, for example charge and spin pumping in
quantum dots~\cite{Switkes1999,Mucciolo2002,Sharma2003,Watson2003},
the role of electron-electron
interactions~\cite{Splettstoesser2005,Sela2006,Reckermann2010},
quantum pumping in graphene mono- and bilayers~\cite{Prada2009,
  Zhu2009, Prada2010, Wakker2010, Tiwari2010, AlosPalop2011,
  Kundu2011} as well as charge and spin pumping through edge states in
quantum Hall systems~\cite{miriam2003} and recently a two-dimensional
topological insulator~\cite{Citro2011}.  On the experimental side,
Giazotto \textit{et al.}~\cite{Giazotto2011} have recently reported an
experimental demonstration of charge pumping in an InAs nanowire
embedded in a superconducting quantum interference device (SQUID).
  
Our main focus is to study quantum pumping induced by periodic
modulations of gate voltages or exchange fields, which are induced by
a ferromagnetic strip in two topological insulator devices: a NFN and
a FNF junction, see Figs.~\ref{fig:NFNjunction}
and~\ref{fig:FNFjunction}. Using a scattering matrix approach, we
obtain analytical expressions for the angle-dependent pumped current
in both types of junctions. We find that the adiabatically pumped
current in a NFN topological insulator junction induced by periodic
modulations of gate voltages reaches maximum values at specific energy
values.  In order to explain the position of these values, we study in
detail the conductance of the junctions. In particular, we provide an
explanation for resonances in the conductance that were predicted but
not explained in detail in the previous
works~\cite{Mondal2010a,Mondal2010b}. We show that each time a new
resonant mode appears in the junction the conductance increases and
the pumped current reaches a maximum value. For the FNF pump we
predict a non-zero current by periodic modulation of the exchange
magnetic coupling in the absence of external voltages. We observe and
analyze basic similarities and differences between the two pumps
studied in this paper and highlight an experimentally distinguishable
feature between the pumped current and the conductance.

The remainder of the paper is organized as follows. In
Sec.~\ref{sec:NFN&FNF}, we describe the NFN and FNF junctions and use
a scattering matrix model to calculate the reflection and transmission
coefficients of both junctions. In Sec.~\ref{sec:conductance}, we
review the conductance of the NFN junction and present a detailed
analysis of the plateau-like steps that appear in the conductance. We
also analyze and compare the conductance of the FNF junction with
parallel and anti-parallel configuration of the magnetization. In
Sec.~\ref{sec:pumpedcurrent}, we calculate the adiabatically pumped
current for the two different pumps and derive analytical expressions
as a function of the angle of incidence of the carriers. We also
investigate the dependence of the pumped current and the conductance
on the width $d$ of the middle region. Finally, in
Sec.~\ref{sec:summary} we summarize our main results and propose
possibilities for experimental observation of our predictions.

\section{NFN and FNF junctions}
\label{sec:NFN&FNF}

\begin{figure}
\includegraphics[width=.8\columnwidth]{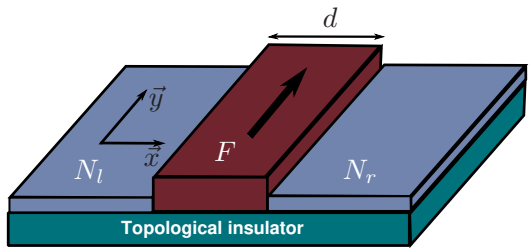}
\caption{(Color online) Sketch of the N$_l$FN$_r$ junction on the
  surface of a topological insulator. Pumping is induced by applying
  gate voltages (not shown) to the normal leads. In the middle region
  a thin ferromagnetic film induces ferromagnetism on the surface of
  the topological insulator by means of the exchange
  coupling~\cite{Yokoyama2010}. The arrow in the middle region
  indicates the direction of the magnetization $M$ in this region.}
\label{fig:NFNjunction}
\end{figure} 
We first describe the NFN junction, see
Fig.~\ref{fig:NFNjunction}. The junction is divided into three
regions: region $N_l$ (for $x < 0$), region $N_r$ (for $x > d$) and
the ferromagnetic region $F$ in the middle.  The left and right-hand
side of the junction represent the bare topological insulator. The
charge carriers (surface Dirac fermions) in these regions are
described by the Hamiltonian $\mathcal{H}_0$
[Eq.~(\ref{eq:hamiltonian0})] whose eigenstates are given by
\begin{equation}
\psi_N^{\pm} = \frac{1}{\sqrt{2}}\, \left(
  \begin{array}{c} 
    1 \\ \pm e^{\pm i \alpha}\\
  \end{array} \right) e^{\pm ik_n x} e^{i q y} ,
\label{eq:eigenstates}
\end{equation}
where $+(-)$ labels the wavefunctions traveling from the left (right)
to the right (left) of the junction. The angle of incidence $\alpha$
and the momentum $k_n$ in the $x$-direction are given by: 
\begin{equation}
  \sin(\alpha) = \frac{\hbar v_F q}{|\epsilon + \mu|},
\label{eq:alpha}
\end{equation}
\begin{equation}
k_n = \sqrt{\left(\frac{\epsilon + \mu}{\hbar v_F}\right)^2 - q^2}.
\end{equation}
Here $\epsilon$ represents the energy measured from the Fermi energy
$\eps_F$ and $q$ denotes the momentum in the $y$-direction. In the
normal regions $N_l$ and $N_r$ a dc electrical voltage can be applied
via metallic top gates to tune the chemical potential $\mu$ and
thereby control the number of charge carriers incident on the
junction.  We assume gate voltages to be small compared to the bandgap
for bulk states ($eV_i \ll E_g \sim 1$ eV, $i=l,r$), so that transport
is well described by surface Dirac states~\cite{HZhang2009}.  In this
case, the eigenstates are given by
\begin{equation}
\psi_{N_l}^{\pm} = \frac{1}{\sqrt{2}}\,  \left(
  \begin{array}{c}
    1 \\ 
    \pm e^{\pm i \alpha_l}\\
  \end{array} \right)
e^{\pm i k_{n_l} x }e^{i q y},
\label{eq:psiNl}
\end{equation}
\begin{equation}
\psi_{N_r}^{\pm} = \frac{1}{\sqrt{2}}\,  \left(
  \begin{array}{c}
    1 \\ 
    \pm e^{\pm i \alpha_r}\\
  \end{array} \right)
e^{\pm i k_{n_r} (x-d) }e^{i q y},
\label{eq:psiNr}
\end{equation}
\begin{equation}
\sin(\alpha_i) = \frac{\hbar v_F q}{|\epsilon + \mu- eV_i|},
\label{eq:alphawithV} 
\end{equation}
\begin{equation}
  k_{n_i} = \sqrt{\left(\frac{\epsilon + \mu - eV_i}
      {\hbar v_F}\right)^2 - q^2},
\end{equation}
where the index $i=l,r$ labels the normal sides of the junction.

In the middle region M of the junction ($0 < x < d$), the presence of
the ferromagnetic strip modifies the Hamiltonian by providing an
exchange field. The Hamiltonian that describes the surface states is now
$\mathcal{H} = \mathcal{H}_0 + \mathcal{H}_{\textrm{induced}}$, where
the induced exchange Hamiltonian is given by~\cite{Yokoyama2010,Mondal2010b}
\begin{equation}
  \mathcal{H}_{\textrm{induced}} = \hbar v_F M \sigma_y,
\end{equation}
with the magnetization $\vec{M} = M\hat{y}$. The magnitude 
$M$ depends on the strength of the exchange coupling of the
ferromagnetic film and can be tuned for soft ferromagnetic films by
applying an external magnetic field~\cite{Yokoyama2010}.  The
eigenstates of the full Hamiltonian $\mathcal{H}$ are then given by:
\begin{equation}
\psi_F^{\pm} = \frac{1}{\sqrt{2}}\, \left(
  \begin{array}{c}
    1 \\ 
    \pm e^{\pm i \alpha_m}\\
  \end{array} \right)
e^{\pm i k_m x}e^{i q y} ,
  \label{eq:psiF}
\end{equation}
with
\begin{equation}
\sin(\alpha_m) = \frac{\hbar v_F (q+M)}{|\epsilon + \mu|},
\label{eq:beta}
\end{equation}
and
\begin{equation}
k_m = \sqrt{\left(\frac{\epsilon + \mu}{\hbar v_F}\right)^2 - (q+M)^2}.
\label{eq:k_m}
\end{equation}
From Eq.~(\ref{eq:k_m}) we see that for a given energy there exists a
critical magnetization 
\be
M_c = \pm\, 2 |\eps + \mu |/(\hbar v_F),
\label{eq:criticalM}
\ee 
beyond which for all transverse ($q$) modes the wavefunction changes from
propagating to spatially decaying (evanescent) along the
$x$-direction~\cite{Mondal2010a,Mondal2010b}.

\begin{figure}
\includegraphics[width=.8\columnwidth]{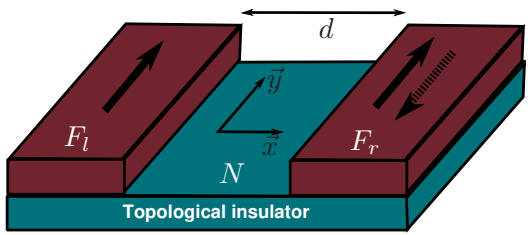}
\caption{(Color online) Sketch of the F$_l$NF$_r$
  junction. Ferromagnetic films are placed on top of the topological
  insulator on the left and right providing exchange fields in these
  regions.  The arrows indicate the direction of the corresponding
  magnetizations $M_l$ and $M_r$, see the text for further
  explanation.}
\label{fig:FNFjunction}
\end{figure} 
Now we describe the FNF junction, see Fig.~\ref{fig:FNFjunction}.
Region $F_l$ ($ x < 0$) and region $F_r$ ($ x > d $) are modeled as
ferromagnetic regions, respectively, with different magnetizations
$M_l$, $M_r$ along the $y$-axis and corresponding wavefunction
$\psi_F$ [Eq.~(\ref{eq:psiF})].  The Dirac fermions in the middle
region N ($ 0 < x < d$) are described by the wavefunctions $\psi_N$
[Eq.(\ref{eq:eigenstates})]. When calculating transport properties of
the FNF junction, we focus on two different alignments of the
magnetizations of the ferromagnetic regions: the parallel
configuration (M$_l$ $\parallel$ M$_r$), where the magnetizations in
the ferromagnetic regions point in the same direction, and the
anti-parallel configuration (M$_l$ $\parallel$ - M$_r$), in which the
magnetizations are in opposite directions.

Using Eqns.~(\ref{eq:eigenstates})-(\ref{eq:k_m}) we can calculate the
reflection and transmission coefficients for a Dirac fermion with
energy $\epsilon$ and transverse momentum $q$ incident from the left
on the junction, for both the NFN and the FNF junctions. To this end,
we consider a general F$_l$F$_m$F$_r$ junction, where the
wavefunctions in each of the three regions left ($l$), middle ($m$)
and right ($r$) are given by:
\begin{eqnarray}
\psi_{l} &=& \psi_{l}^+ + r_{ll}\, \psi_{l}^- ,
\nonumber \\ 
\psi_m &=& p\, \psi_m^+ + q\, \psi_m^- ,
\\ 
\psi_{r} &=& t_{rl}\, \psi_{r}^+.
\nonumber
\end{eqnarray}
Here $\psi_{j}^{\pm}$ ($j = l, m ,r$) are the wavefunctions
(\ref{eq:psiNl}), (\ref{eq:psiNr}) or (\ref{eq:psiF}) (depending on
the junction considered) and $r_{ll}$ and $t_{rl}$ denote the
corresponding reflection and transmission coefficients. By requiring
continuity of the wavefunction at the interfaces $x = 0$ and $x = d$,
we obtain the reflection and transmission coefficients:
\begin{widetext}
\begin{eqnarray}
r_{l l} &=& e^{i \alpha_l}\, \frac{e^{2 i k_m d} (1 + e^{i (\alpha_m+
    \alpha_l)}) (e^{i \alpha_m} - e^{i \alpha_r}) + (e^{ i \alpha_l} -
  e^{i \alpha_m}) (1 + e^{i (\alpha_m+ \alpha_r)})}{e^{2 i k_m d}
  (e^{i \alpha_m} - e^{i \alpha_l}) (e^{i \alpha_m} - e^{i \alpha_r})
  + (1 + e^{i (\alpha_m+ \alpha_l)}) (1 + e^{i (\alpha_m+ \alpha_r)})}
,
  \label{eq:generalreflectionLL}\\
t_{rl} &=& \frac{e^{i k_m d}(1 + e^{2 i \alpha_m}) (1 + e^{2 i \alpha_l})} {e^{2
    i k_m d} (e^{i \alpha_m} - e^{i \alpha_l}) (e^{i \alpha_m} - e^{i
    \alpha_r}) + (1 + e^{i (\alpha_m+ \alpha_l)}) (1 + e^{i (\alpha_m+
    \alpha_r)})} .
\label{eq:generaltransmissionLR}
\end{eqnarray} 
\end{widetext}
Here $\alpha_j$ denotes the polar angle of the wavevector in region
$j=l,m,r$ [Eqns.~(\ref{eq:alphawithV}) and (\ref{eq:beta})]. When
considering an electron incident from the right lead, one can
similarly obtain $r_{rr}$ and $t_{lr}$.
These expressions for the reflection and transmission coefficients
form the basis of our calculations of the conductance and the pumped
current in Secs.~\ref{sec:conductance} and \ref{sec:pumpedcurrent}
respectively.

\section{Conductance}
\label{sec:conductance}

\begin{figure}
\includegraphics[width=.8\columnwidth]{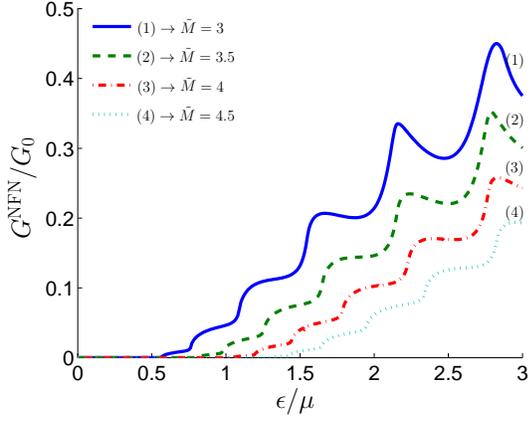}
\caption{(Color online) The conductance $G^{\text{NFN}}$ of the NFN junction
  [Eq.(\ref{eq:conductance})] as a function of $\epsilon / \mu$ for
  $V_l=V_r=0$ and for different values of the effective magnetization
  $\tilde{M}\equiv \hbar v_F M / \mu = 3$ (solid blue line), 3.5
  (dashed green line), 4 (dot-dashed red line) and 4.5 (dotted
  light-blue line). The effective junction width $\tilde{d} \equiv \mu
  d/(\hbar v_F) = 5$.}
\label{fig:NFNconductance}
\end{figure}

The conductance $G^{\text{NFN}}$ of a topological insulator NFN
junction has been studied in earlier work by Mondal \textit{et
  al.}~\cite{Mondal2010a,Mondal2010b}, who predicted 
oscillatory behavior of $G^{\text{NFN}}$ as a function of the applied
bias voltage (see also Fig.~\ref{fig:NFNconductance}). In this section
we first briefly review their results and then add a quantitative
explanation for the oscillations of the conductance. This explanation
is crucial for understanding the behavior of the pumped current in the
next section. We also calculate and analyze the conductance in a FNF
junction.

The general expression for the conductance $G$ across the junction in
terms of the transmission probability $T(\al) \equiv |t_{rl}(\al)|^2$
is given by
\begin{equation}
G = (G_0 /2) \int_{-\pi/2}^{\pi/2} T(\al) \cos \alpha\, d\alpha.
\label{eq:conductance}
\end{equation}
Here $G_0= \frac{2e^2}{h} \rho(eV)\hbar v_F W $, $\rho(eV)= | \mu+eV |
/(2\pi (\hbar v_F)^2)$ denotes the density of states, $W$ is the
sample width, and the integration is over all the angles of incidence
$\al$. For $V_l=V_r=0$ ($\alpha_l=\alpha_r=\alpha$) the
angle-dependent transmission probability $T^{\text{NFN}}(\al)$ is
given by~\cite{Mondal2010a,Mondal2010b}
\begin{eqnarray}
T^{\text{NFN}}(\al) &=&\cos^2(\alpha)\cos^2(\alpha_m)/ \left[
  \cos^2(k_md)\cos^2(\alpha) \cos^2(\alpha_m) \right. \nonumber \\
&&\left. + \sin^2(k_md)(1-\sin(\alpha)\sin(\alpha_m))^2 \right],
\label{eq:totaltransmission}
\end{eqnarray}
where $\alpha_m$ is the polar angle of the wave vector in the middle
region as defined in Eq.~(\ref{eq:beta}). This angle can be expressed
in terms of $\alpha$ using the fact that the momentum is conserved
along $y$-axis as:
\begin{equation}
\sin(\alpha_m) = \sin(\alpha) + \frac{M\hbar v_F}{\mid\epsilon+\mu\mid}.
\label{eq:alpham}
\end{equation}
Figure~\ref{fig:NFNconductance} shows the conductance of the NFN
junction [obtained from Eqns.~(\ref{eq:conductance}) and
  (\ref{eq:totaltransmission})] as a function of the energy of the
incoming carriers $\epsilon / \mu$ for different values of the
effective magnetization $\tilde{M} \equiv \hbar v_F M / \mu$. For a
given magnetization $M$, the conductance is zero for $\epsilon <
\epsilon_c$, i.e., $G^{\text{NFN}}(\epsilon)=0$, with the critical
energy $\epsilon_c \equiv \hbar v_F M / 2 - \mu$. Below this energy
there are no traveling modes inside the barrier. Our results agree
with the previous results in the
literature~\cite{Mondal2010a,Mondal2010b}.

From Fig.~\ref{fig:NFNconductance} it can be observed that the
conductance changes from plateau-like to oscillatory as $\epsilon/\mu$
increases. In order to provide an explanation for this behavior we
first analyze the plateau like regime in detail. After setting
$\alpha_l = \alpha_r \equiv \alpha$ in
Eq.~(\ref{eq:generalreflectionLL}), we begin by finding the conditions
when the reflection coefficient is zero, i.e., $r_{ll}=0$. The first,
trivial, condition $\alpha = \alpha_m + 2\pi n$ corresponds to the
situation of an entirely normal junction (i.e., no ferromagnetic
region). The second and more interesting condition is $\sin (k_m d) =
0$. This is the case when transmission occurs via a \textit{resonant}
mode of the junction and can be written as (using
Eqns.~(\ref{eq:alpha}) and (\ref{eq:k_m}))
\begin{equation}
k_m d = \frac{|\epsilon + \mu|}{\hbar v_F}d\sqrt{1 - \left( \sin
  \alpha + \frac{\hbar v_F M}{|\epsilon + \mu|} \right)^2}=n\pi.
\label{eq:resonantcondition}
\end{equation}
Eq.~(\ref{eq:resonantcondition}) indicates that for a given $M$ and
$\epsilon$ there are certain privileged angles $\al_c$ for which the
barrier becomes transparent:

\begin{equation}
 \sin (\alpha_c) = \pm \sqrt{1 - \left( \frac{n\pi}{\tilde{d} (1 +
     \frac{\eps}{\mu})} \right)^2} - \frac{\tilde{M}}{|1 +
   \frac{\eps}{\mu}|},
\label{eq:condition}
\end{equation}
with $\tilde{d} \equiv d \mu/(\hbar v_F)$ being the dimensionless barrier
length.  These modes 
are referred to as \textit{resonant} modes in this article.

\begin{figure}%
\subfigure{\includegraphics[width=.48\columnwidth]{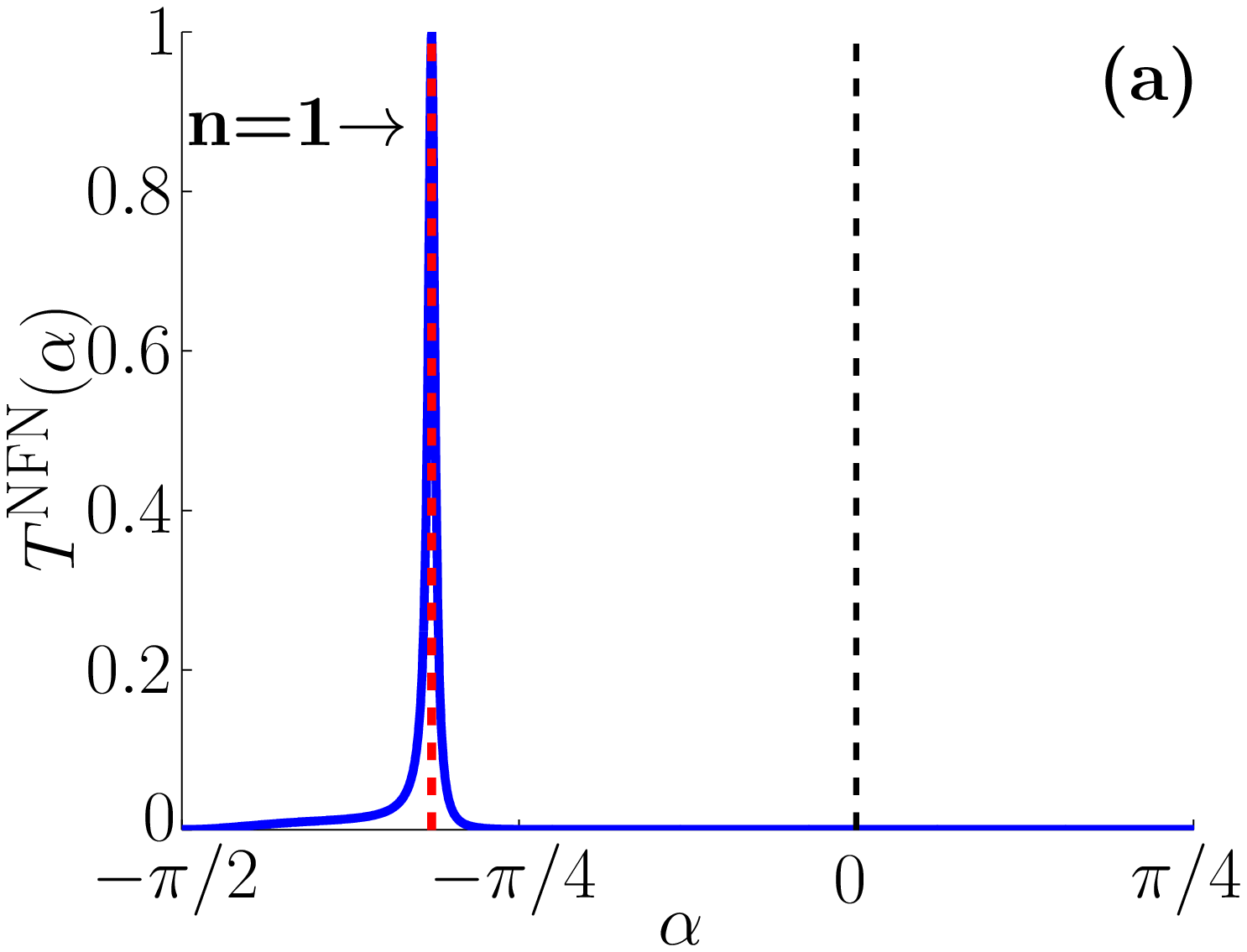}}\hfill
\subfigure{\includegraphics[width=.48\columnwidth]{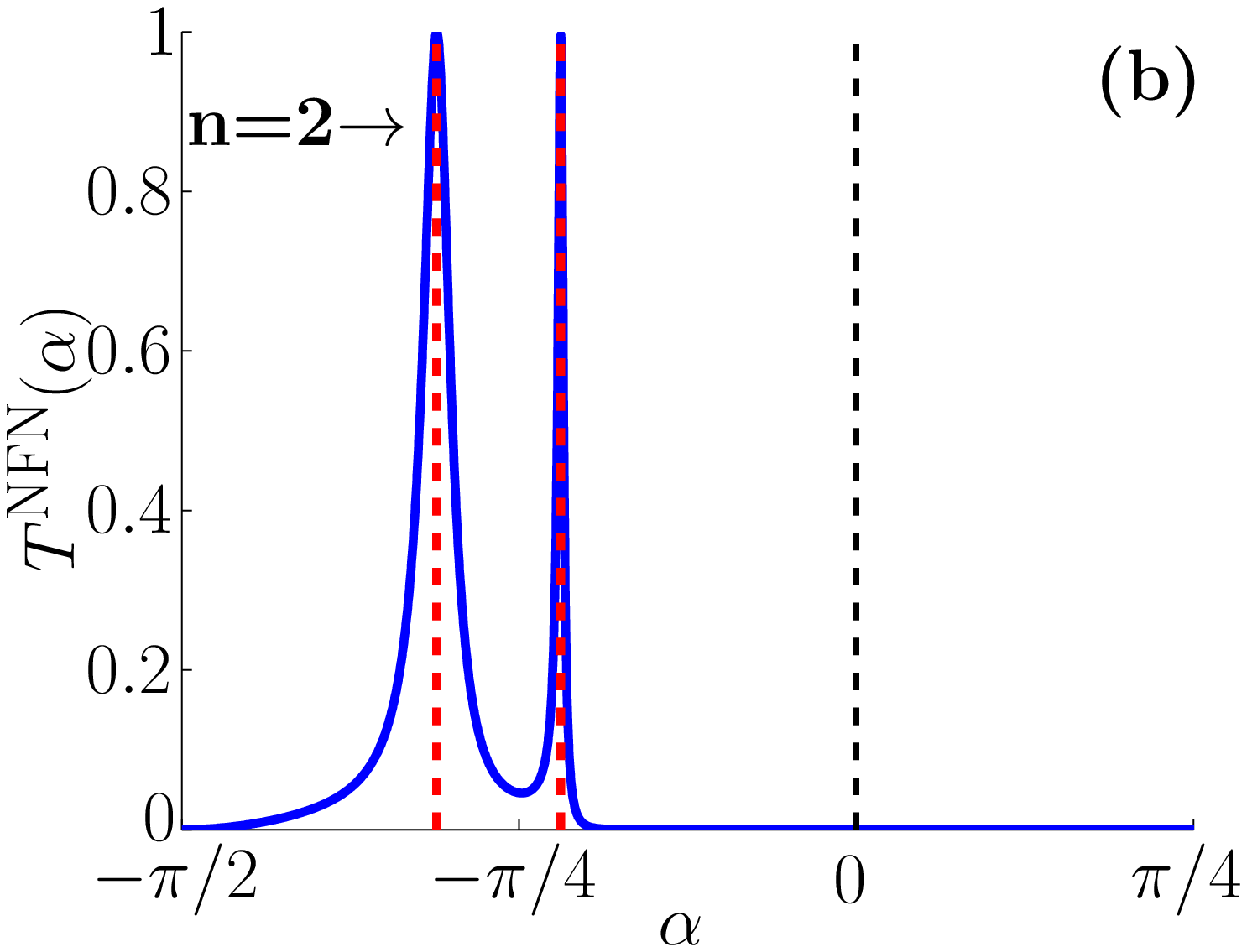}}\\
\subfigure{\includegraphics[width=.48\columnwidth]{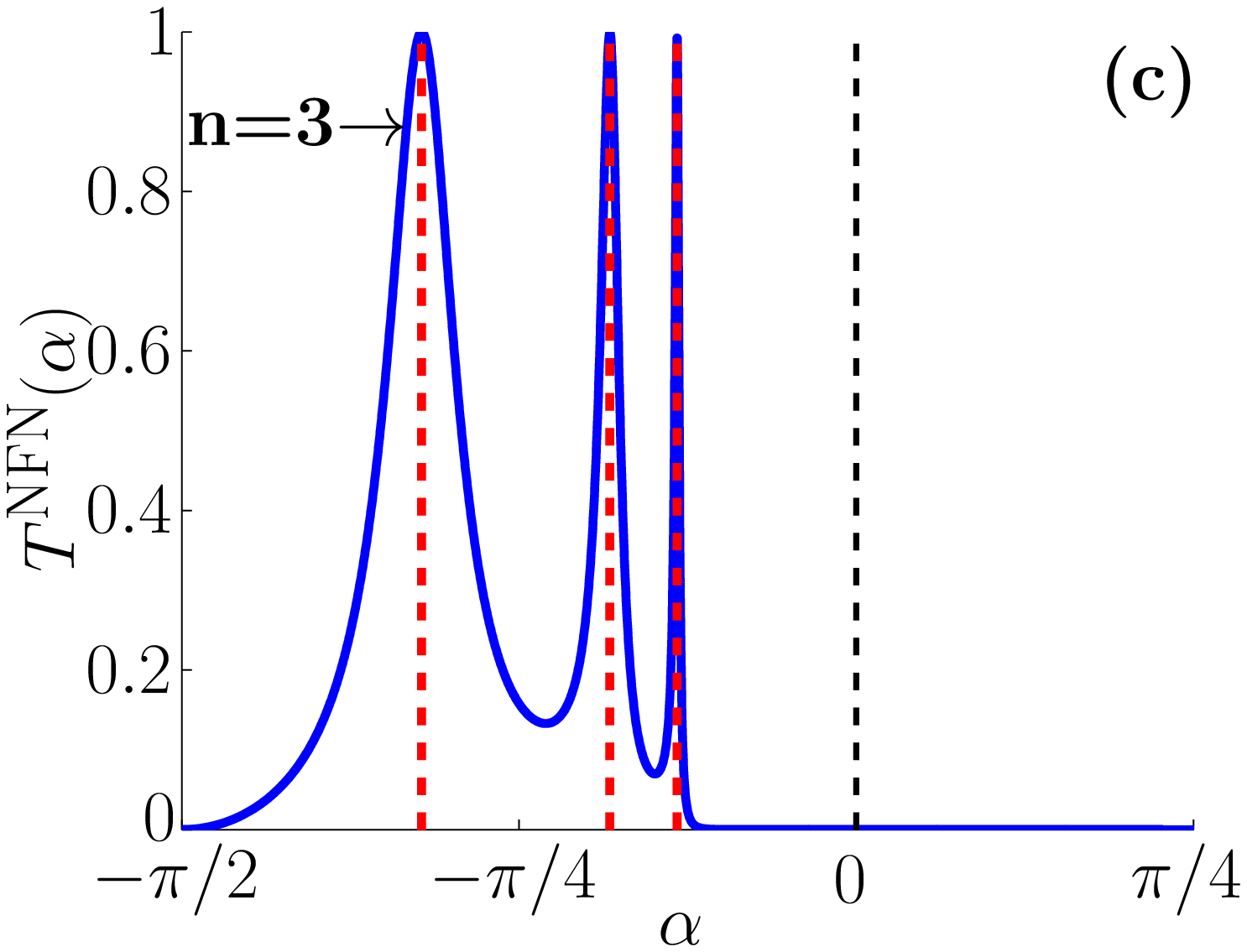}}\hfill
\subfigure{\includegraphics[width=.48\columnwidth]{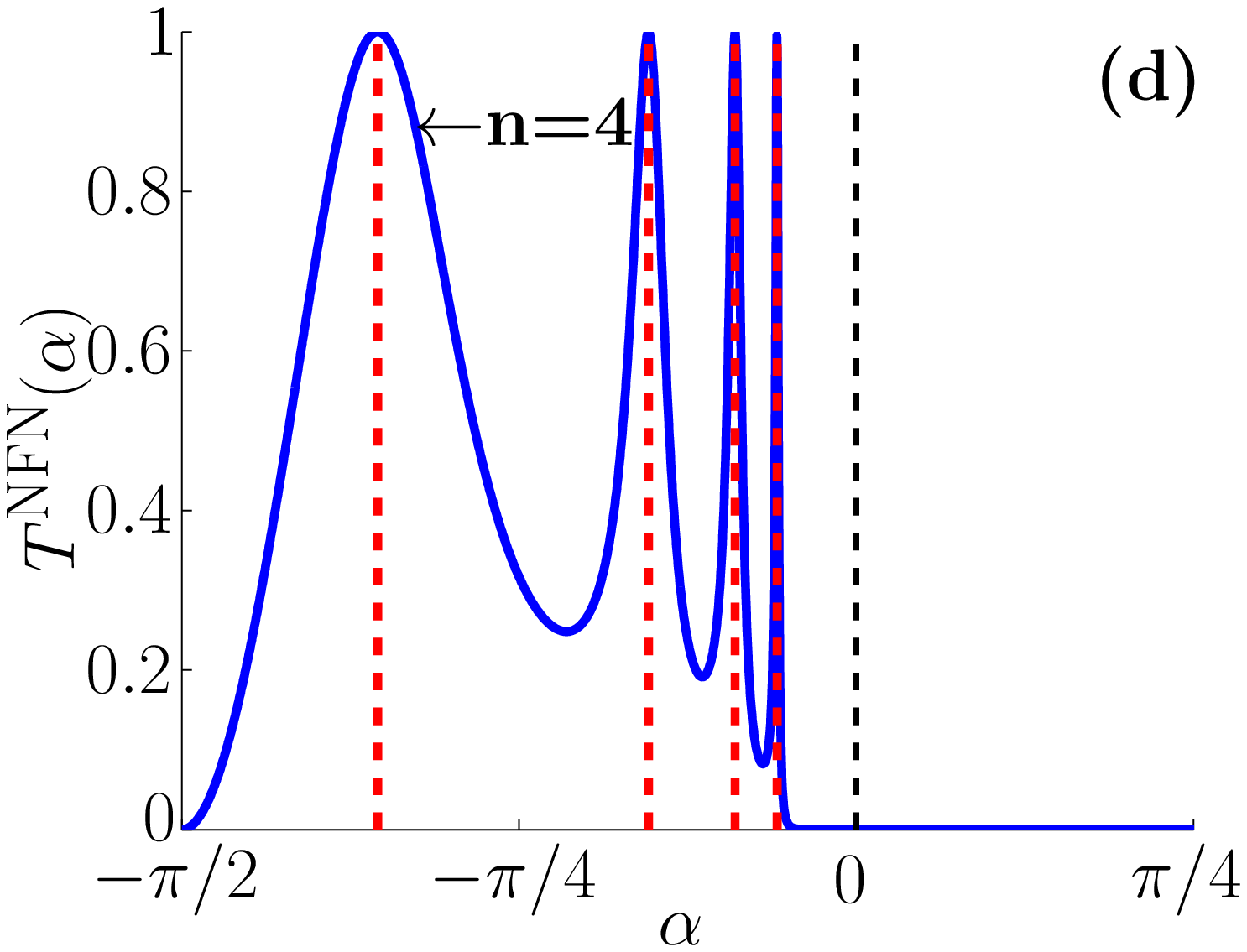}}\\
\subfigure{\includegraphics[width=.48\columnwidth]{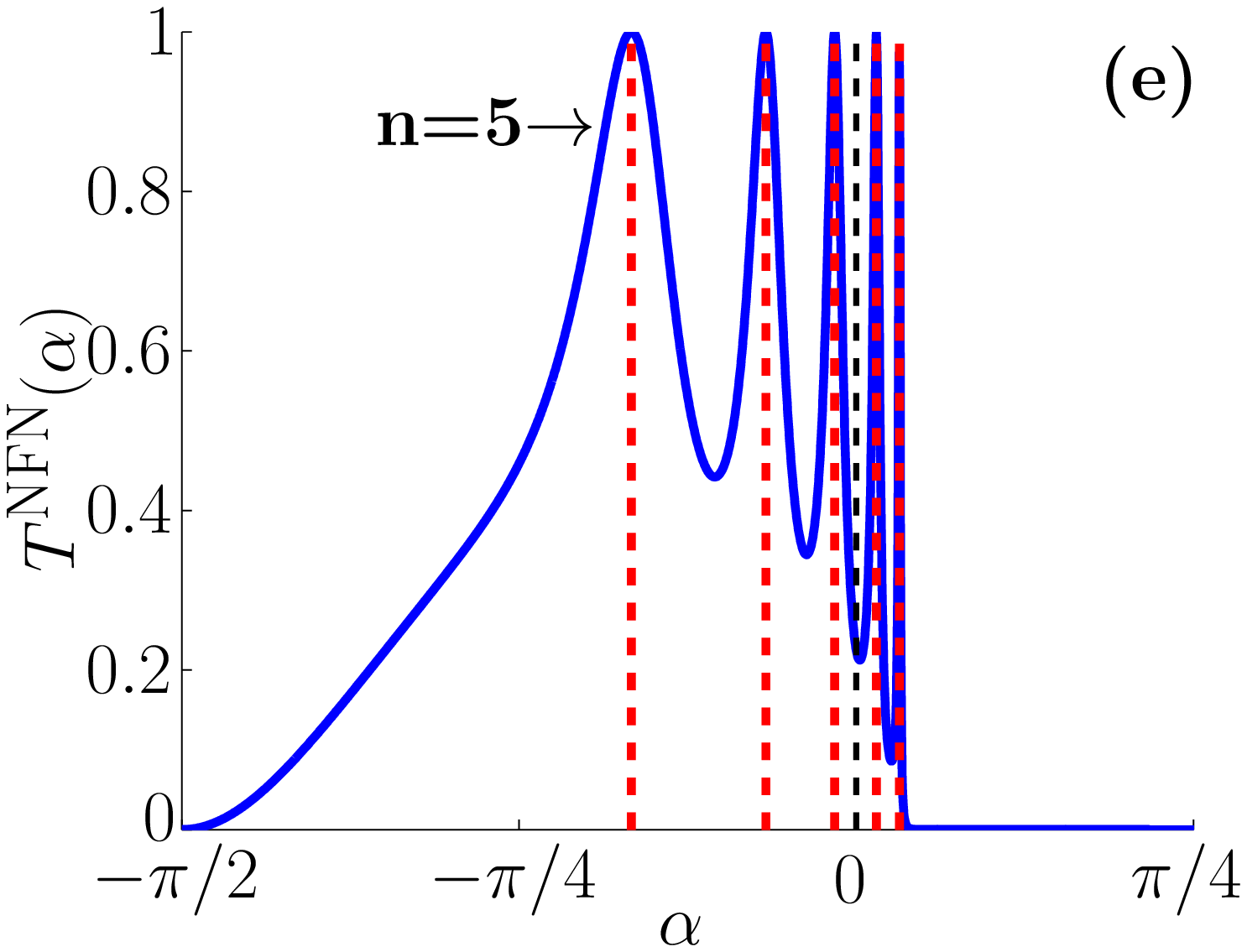}}\hfill
\subfigure{\includegraphics[width=.48\columnwidth]{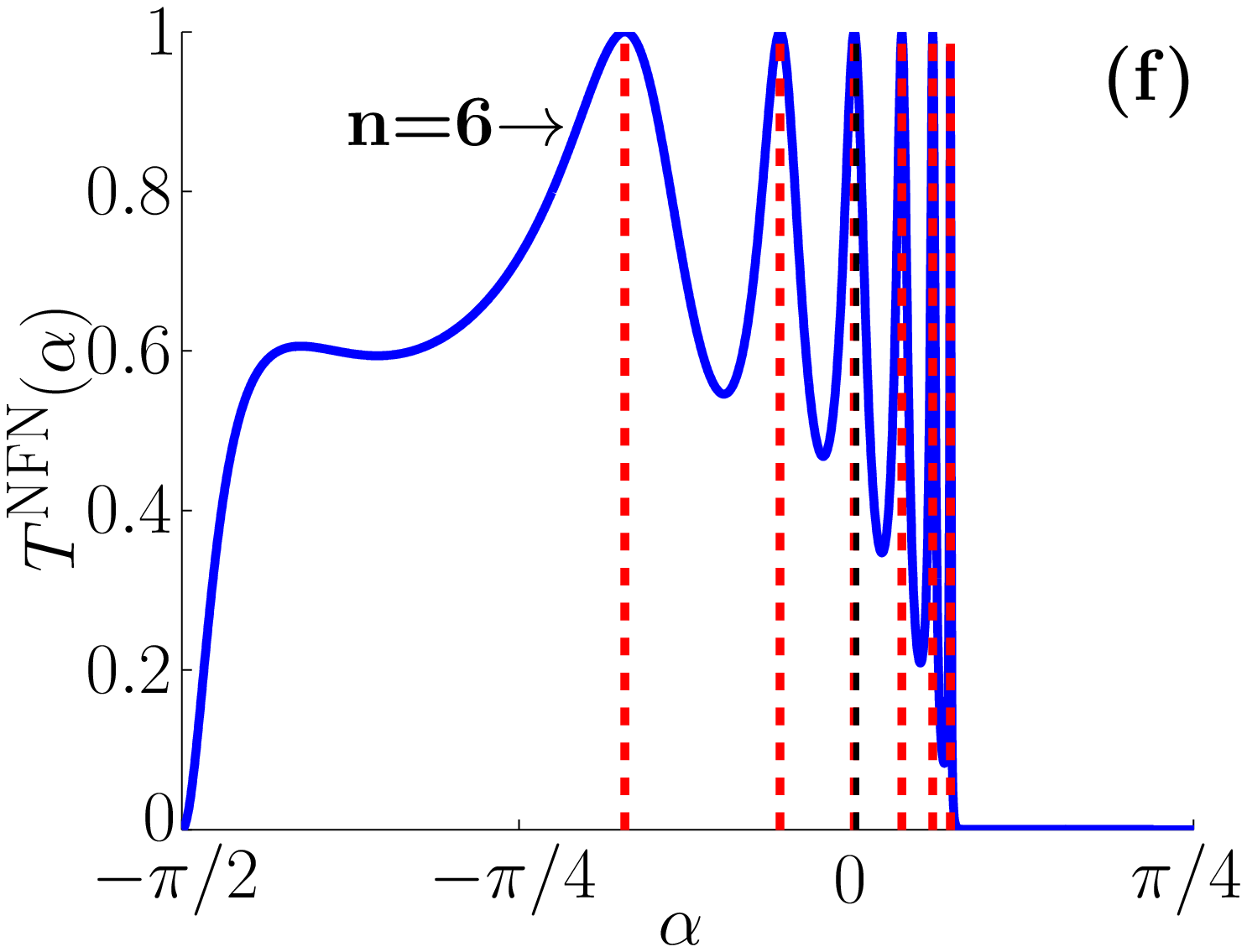}}
\caption{(Color online) The transmission probability $T^{\text{NFN}}(\al)$
  [Eq.~(\ref{eq:totaltransmission})] as a function of the angle of
  incidence $\alpha$ for different values of energy $\epsilon/\mu$,
  (a) $\epsilon/\mu = 0.7$, (b) $\epsilon/\mu = 0.9$, (c)
  $\epsilon/\mu = 1.2$, (d) $\epsilon/\mu = 1.6$, (e)$\epsilon/\mu =
  2.4$, and (f) $\epsilon/\mu = 2.9$.  Parameters used are $\tilde{d}
  = 5$ and $\tilde{M} = 3$.}
\label{fig:Gvsalpha}
\end{figure}
Figure~\ref{fig:Gvsalpha} shows the transmission probability
[Eq.~(\ref{eq:totaltransmission})] as a function of the angle of
incidence $\alpha$ for different values of energy $\epsilon/\mu$. The
dashed (red) vertical lines correspond to the angles satisfying
Eq.~(\ref{eq:condition}) for different $n$. It can be seen that as the
energy increases more resonant modes become available for
transmission. It is also worth noting that for energies at which only
one mode is present ($n=1$, see Fig.~\ref{fig:Gvsalpha}(a)) the
transmission is strongly localized at one particular angle. This
property could be exploited to fabricate single-mode filters. For low
energy excitations only negative angles $\alpha$ (i.e., $q$-momenta
anti-parallel to M) contribute to the conductance, see
Figs.~\ref{fig:Gvsalpha}(a)-(d). As the energy increases, the resonant
modes move from the left to the right and also positive angles
$\alpha$ (i.e., $q$-momenta parallel to M) begin to contribute, see
Figs.~\ref{fig:Gvsalpha}(e) and (f).

\begin{figure}
\includegraphics[width=.8\columnwidth]{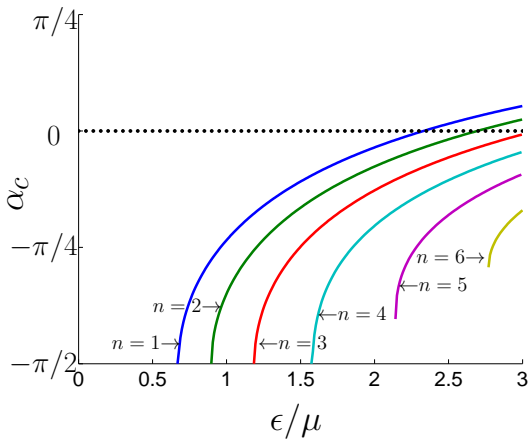}
\caption{(Color online) The real part of the angle of incidence $\alpha_c$
  [Eq.~(\ref{eq:condition})] versus energy $\eps/\mu$ for the modes
  $n=1,2,3,4,5$. Parameters used are the same as in
  Fig.~\ref{fig:Gvsalpha}.}
\label{fig:alphavsen}
\end{figure}

Now we address the question why a mode becomes resonant in the
barrier. Figure~\ref{fig:alphavsen} shows the angle of incidence
$\alpha_c$ [Eq.~(\ref{eq:condition})] for different values of $n$ as a
function of energy $\epsilon/\mu$. For a given $n$, the resonant mode
does not contribute to the conductance if the energy $\epsilon$
satisfies the condition $ \epsilon_c < \epsilon <
\epsilon_c^{\alpha_n}$, because the imaginary part of the momentum
$k_m$ is nonzero and thus the mode is decaying along the
$x$-direction. This critical energy $\eps_c^{\alpha_n}$ for each mode
is given by:
\begin{equation}
\frac{\epsilon_c^{\alpha_n}}{\mu}= \frac{n^2
  \pi^2}{2\tilde{d}^2 \tilde{M}} + \frac{\tilde{M}}{2} -1.
 \label{eq:EnResonantMode}
\end{equation}

When $\epsilon > \epsilon_c^{\alpha_n}$, $k_m$ becomes real
and the mode becomes resonant. As we increase
the energy, all the modes asymptotically reach their saturation angle
$\alpha = \pi/2$.

Finally, we analyze the effect of the appearance of subsequent
resonant modes on the conductance. 
For small energies, the conductance increases in plateau-like steps
(see Fig.~\ref{fig:NFNconductance}). The first plateau corresponds to
the situation in which the first transmission mode appears at
$\alpha=-\pi/2$. As the energy increases, a new resonant mode appears
and the conductance increases in a step-like manner. The plateaus are
not sharp due to the fact that each new mode appearing is not sharply
peaked, but rather has a certain distribution around a particular
angle of incidence, see Fig.~\ref{fig:Gvsalpha}. Once the energy is
large enough for there to be contributions from both positive and
negative angles of incidence, the conductance becomes oscillatory. For
very large energies ($\epsilon \gg \epsilon_c$), the effect of the
magnetic barrier disappears and the conductance becomes unity
($G^{\text{NFN}}= G_0$).

\begin{figure}[tb]
\includegraphics[width=.8\columnwidth]{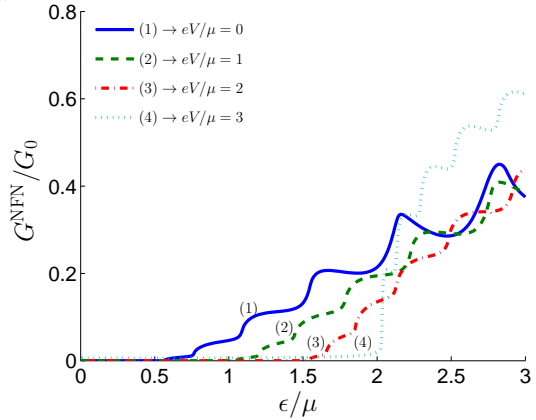}
\caption{(Color online) The conductance $G^{\text{NFN}}$ of the NFN
  junction as a function of $\epsilon / \mu$ for different values of
  gate voltages, $eV/\mu=0$ (solid blue line), 1 (dashed green line),
  2 (dashed-dot red line) and 3 (double-dotted light-blue line). As
  before, $\tilde{M} = 3$ and $\tilde{d} =5$.}
\label{fig:N1MN2conductanceOverV}
\end{figure}

Figure~\ref{fig:N1MN2conductanceOverV} shows the conductance as a
function of energy for several values of applied bias voltages $V_l =
V_r \equiv V$. As expected, the features of the conductance remain the
same for finite $V$. As we increase $eV$, the critical energy
$\epsilon_c = \hbar v_F M/2 +eV/2 -\mu$ for the onset of the
conductance increases and the spacing between two consecutive resonant
modes decreases. As a result the plateaus become narrower.

In the remaining part of this section we study the conductance in a
topological insulator FNF junction, as shown in
Fig.~\ref{fig:FNFjunction}. We consider both the junction with
parallel and with anti-parallel magnetization in the ferromagnetic
regions. In the parallel configuration, using $\alpha_l = \alpha_r
\equiv \alpha = \sin^{-1} ( \hbar v_F (q+M)/ |\epsilon + \mu| )$ and
$\alpha_{m} = \sin^{-1} ( \hbar v_F q/|\epsilon + \mu|)$ in
Eq.~(\ref{eq:generaltransmissionLR}), we find that the conductance is
similar to the conductance of a NFN junction, as displayed in
Fig.~\ref{fig:NFNconductance}.  However, in the FNF junction the first
resonant mode becomes resonant for positive $\alpha$ (i.e., transverse
$q$-momentum parallel to $M$) and as the energy increases, the
resonances move towards negative values of the angle $\alpha$. This,
however, does not affect the total conductance, as we sum over all
possible angles of incidence, and the same analysis as for the NFN
junction presented above can be applied to understand the FNF junction
with parallel magnetization.

In the case of anti-parallel alignment of the magnetization in the two
ferromagnetic regions, we substitute $\sin(\alpha_l) = \hbar v_F (q+M)/
|\epsilon + \mu|$, $\sin(\alpha_r) = \hbar v_F (q-M)/ |\epsilon + \mu|$
and $\sin(\alpha_{m}) =\hbar v_F q/|\epsilon + \mu|$ in
Eq.~(\ref{eq:generaltransmissionLR}). The conductance of this junction
was studied previously in Refs.~\cite{Wu2010,Salehi2011} and the
transmission probability $T^{\text{FNF,AP}}(\al_l,\al_r) \equiv 
|t_{rl}(\al_l,\al_r)|^2 $ is given by:
\begin{widetext}
\be
T^{\text{FNF,AP}}(\al_l,\al_r)=
\frac{\cos^2(\al_l)  \cos^2(\al_m)}
{\cos^2(k_md) \cos^2(\frac{\al_l + \al_r}{2}) \cos^2(\al_m) +
\sin^2(k_md) \left[ \cos \left( \frac{\al_l - \al_r}{2} \right) -
  \sin(\frac{\al_l + \al_r}{2}) \sin(\alpha_m) \right]^2}.
\label{eq:antiparallel}
\ee
\end{widetext}
The total conductance is obtained by multiplying
$T^{\text{FNF,AP}}(\al_l,\al_r)$ with $\cos(\alpha_r)/\cos(\alpha_l)$
and then integrating over the allowed angles of
incidence~\cite{restriction}, i.e., from $\alpha_{c_1} = \sin^{-1} (2
\hbar v_F M/(|\epsilon + \mu|) - 1)$ to $\alpha_{c_2} =\sin^{-1} (2
\hbar v_F M/(|\epsilon + \mu|) +1)$. Thus we can write
\begin{equation}
G^{\text{FNF,AP}}=G_0/2\int_{\al_{c_1}}^{\pi/2}G^{\text{FNF,AP}}
(\al_l,\al_r)\cos(\al_l)d\al_l,
\label{eq:newequation}
\end{equation}
where
\begin{equation}
G^{\text{FNF,AP}}(\al_l,\al_r) = \frac{\cos(\alpha_r)}{\cos(\alpha_l)}
T^{\text{FNF,AP}}(\al_l,\al_r).
\label{eq:neweq2}
\end{equation}

\begin{figure}
\includegraphics[width=.8\columnwidth]{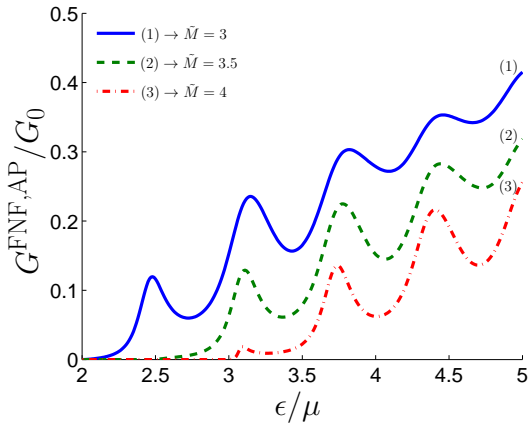}
\caption{(Color online) The conductance $G^{\text{FNF,AP}}$ of the FNF junction in
  the anti-parallel configuration as a function of $\epsilon /\mu$,
  for $\tilde{M} = 3$ (blue solid line), $3.5$ (green dashed line) and
  $4$ (red dotted line). The parameter $\tilde{d}=5$.}
\label{fig:FNFantiparallelVSparallel}
\end{figure}
\begin{figure}
\subfigure{\includegraphics[width=.48\columnwidth]{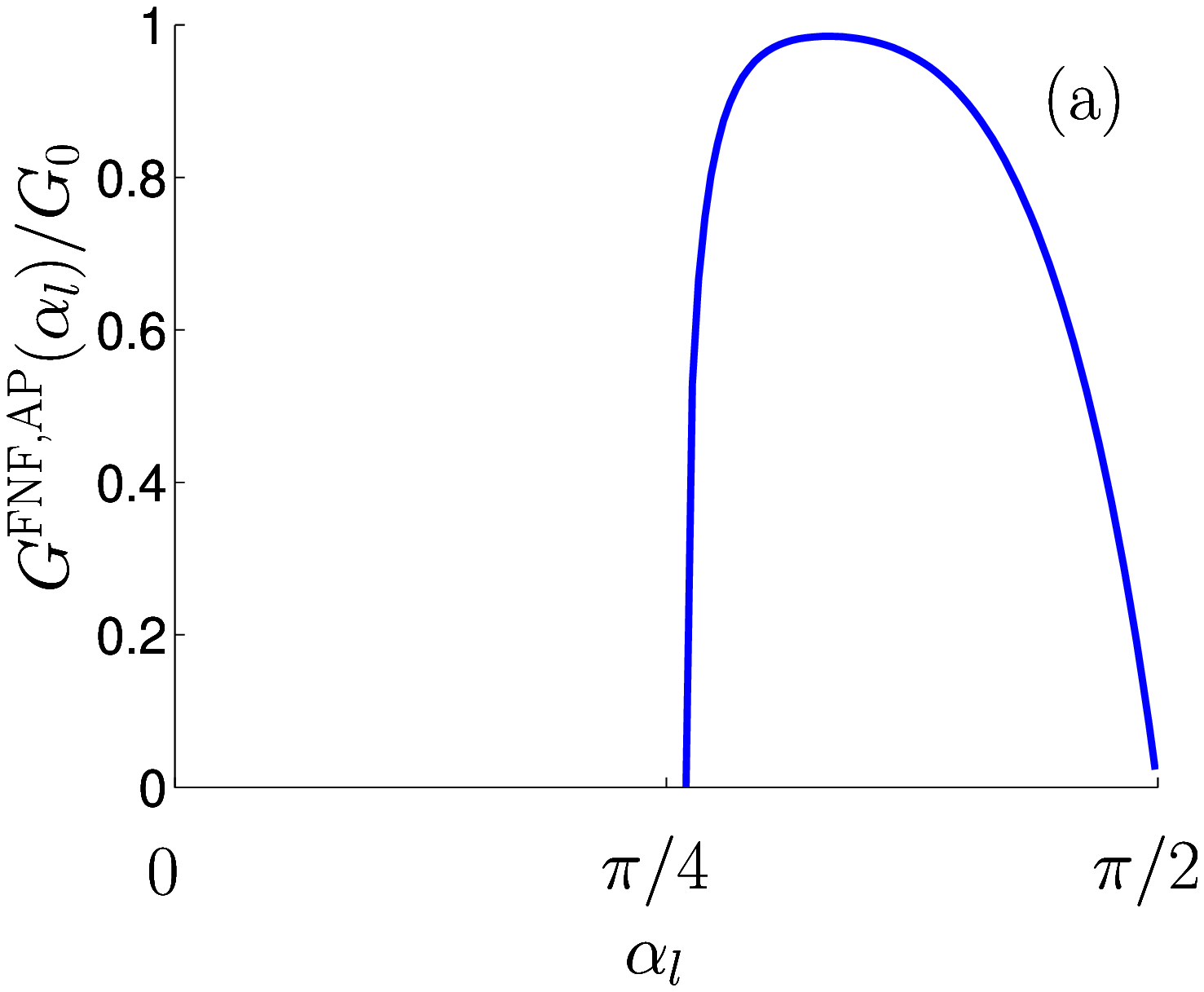}}\hfill
\subfigure{\includegraphics[width=.48\columnwidth]{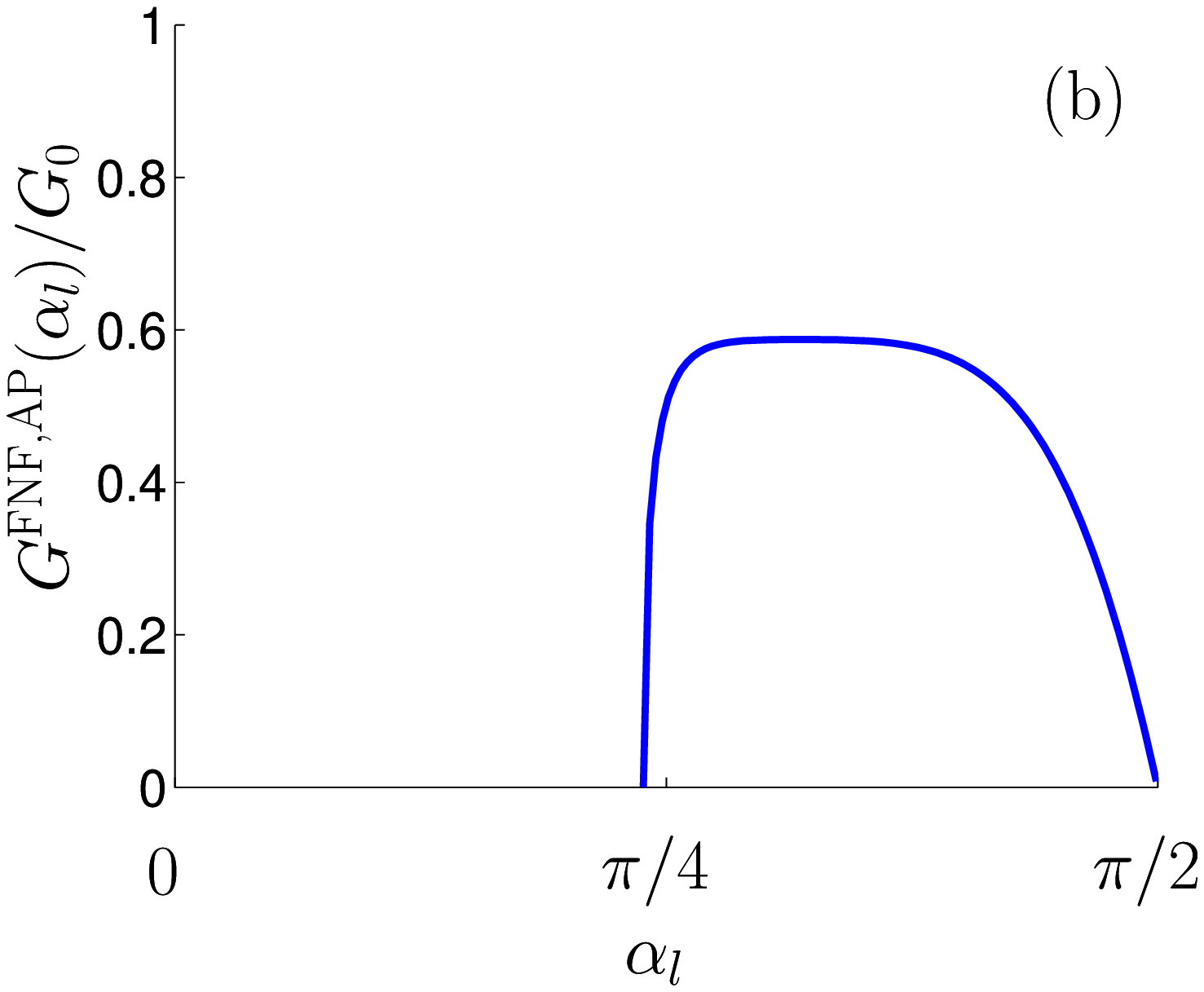}}
\subfigure{\includegraphics[width=.48\columnwidth]{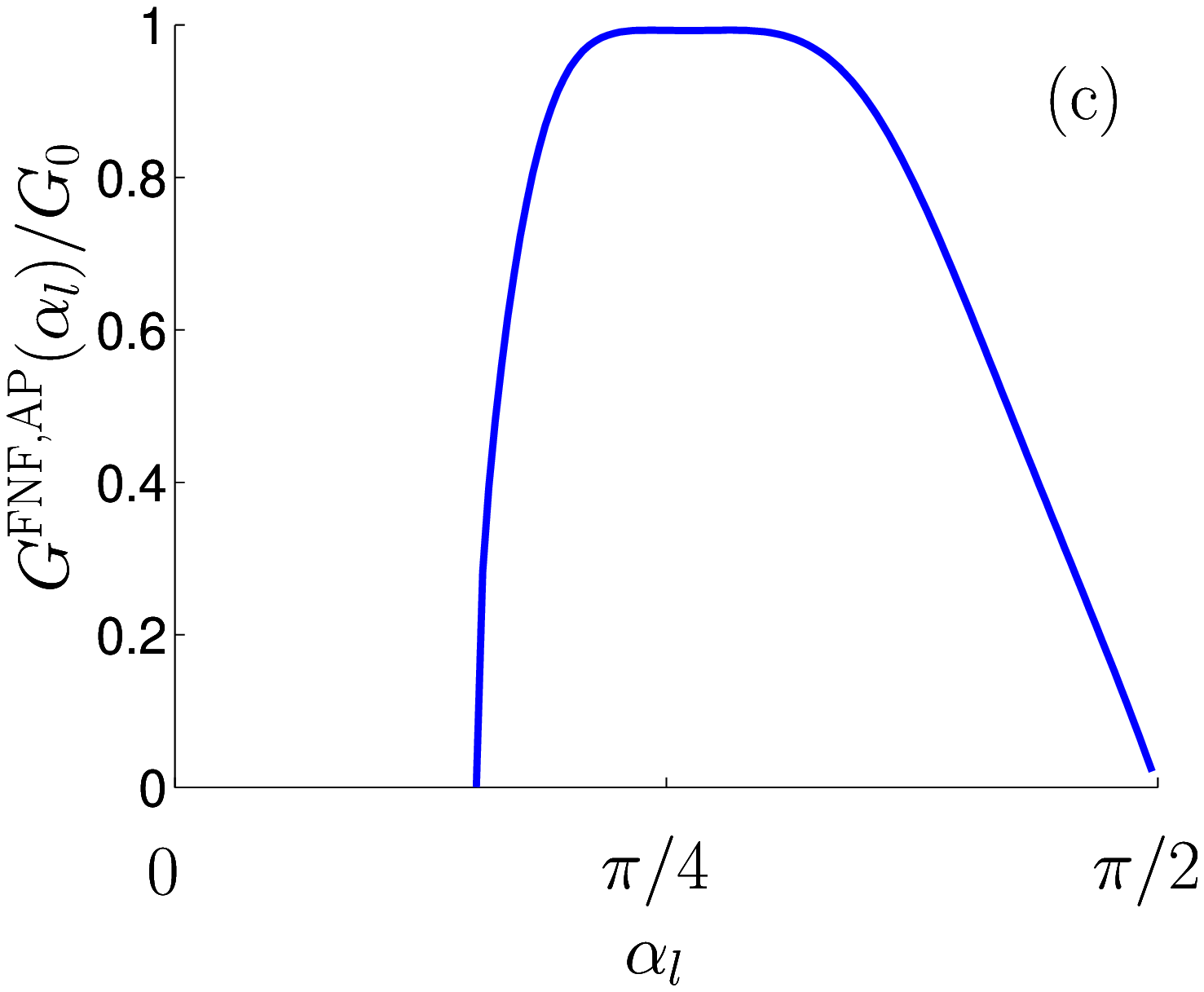}}\hfill
\subfigure{\includegraphics[width=.48\columnwidth]{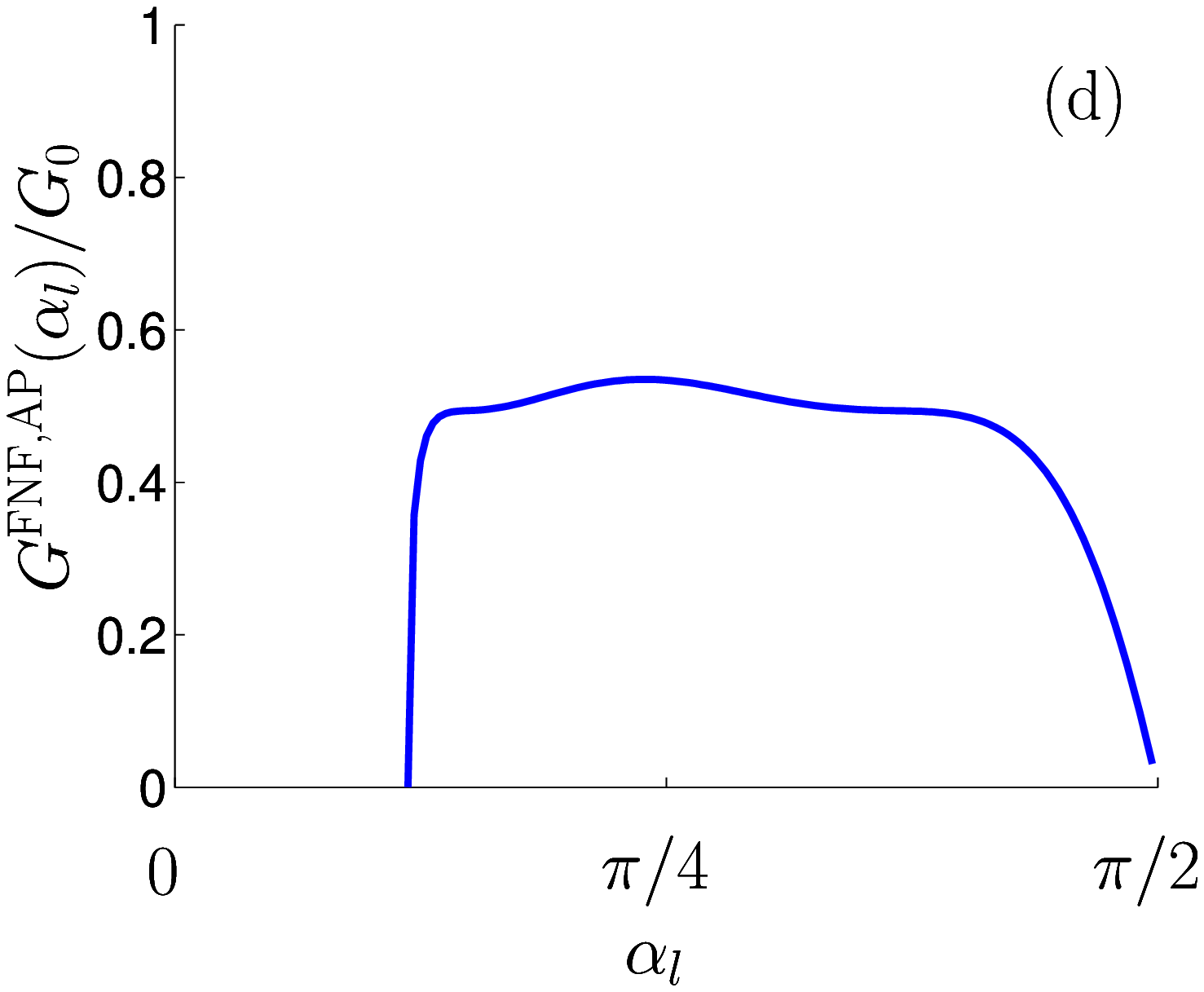}}
  \caption{The angle-dependent total transmission
    $T^{\text{FNF,AP}}(\al_l,\al_r)$ for the anti-parallel
    configuration of the FNF junction as a function of the angle of
    incidence~$\alpha_l$ for (a) $\epsilon/\mu = 2.47$, (b)
    $\epsilon/\mu = 2.57$, (c) $\epsilon/\mu = 3.10$ and (d)
    $\epsilon/\mu = 3.40$. Parameters used are $\tilde{M} = 3$ and
    $\tilde{d} = 5$.}
\label{fig:GANTIvsalpha}
\end{figure}

Fig.~\ref{fig:FNFantiparallelVSparallel} shows the conductance of the
FNF junction in the anti-parallel configuration. From the horizontal
axis we see that the critical energy $\eps_c$ for the onset of the
conductance is larger than in the corresponding parallel
configuration. Moreover, as the energy $\epsilon/\mu$ increases the
conductance exhibits no plateau behavior: it increases in an
oscillatory fashion.  This oscillatory behavior can be understood from
Fig.~\ref{fig:GANTIvsalpha}, which shows $G^{\text{FNF,AP}}(\alpha)$
for four different values of $\epsilon/\mu$. Note that all the angles
($\alpha_l$, $\alpha_m$ and $\alpha_r$) can be expressed in terms of
one angle, which we choose to be $\alpha_l\equiv\alpha$. As the energy
increases, the area under the curve oscillates resulting in
oscillations in the conductance.

Summarizing, we have obtained a quantitative explanation for the
behavior of the conductance in topological insulator NFN and
FNF-junctions in terms of the number of resonant modes in the
junction. This explanation forms the basis for understanding the
behavior of the pumped current in the next section.
\\

\section{Adiabatically pumped current}
\label{sec:pumpedcurrent}

In this section we investigate adiabatically pumped currents through
NFN and FNF junctions in a topological insulator~\cite{difference}.
In general, a pumped current is generated by slow variation of two
system parameters $X_1$ and $X_2$ in the absence of a bias
voltage~\cite{Buttiker1994,Brouwer1998}. For periodic modulations
$X_1(t)= X_{1,0} + \delta X_1 \cos (\omega t)$ and $X_2(t) = X_{2,0} +
\delta X_2 \cos (\omega t + \phi)$, the pumped current $I_p$ into the
left lead of the junction can be expressed in terms of the area $A$
enclosed by the contour that is traced out in $(X_1,X_2)$-parameter
space during one pumping cycle as~\cite{Brouwer1998}:
\begin{subequations}
\begin{eqnarray}
I_p & = & \frac{\omega e}{2 \pi^2} \int_A\, dX_1\, dX_2\,
\sum_{m} \ \Pi (X_1,X_2)
\label{eq:Ip} \\
& \approx & \frac{\omega e}{2 \pi}\, \delta X_1\, \delta X_2\, \sin
\phi \, \sum_{m} \ \Pi (X_1,X_2),
\label{eq:IpLinearResp}
\end{eqnarray}
\end{subequations} 
with 
\begin{equation} 
\Pi (X_1,X_2) \equiv \mbox{\rm Im}\ \left( \frac{\partial
  r_{ll}^{\ast}}{\partial X_1} \frac{\partial r_{ll} }{\partial X_2} +
\frac{\partial t_{lr}^{\ast}}{\partial X_1} \frac{\partial
  t_{lr}}{\partial X_2} \right).
\label{eq:Pifunc}
\end{equation}
Here $r_{ll}$ and $t_{lr}$ represent the reflection and transmission
coefficients into the left lead and the index $m$ sums over all modes (a
similar expression can be obtained for the pumped current into the right
lead). Eq.~(\ref{eq:IpLinearResp}) is valid in the bilinear response
regime where $\delta X_1 \ll X_{1,0}$ and $\delta X_2 \ll X_{2,0}$ and
the integral in Eq.~(\ref{eq:Ip}) becomes independent of the pumping
contour.

\begin{figure}
\subfigure{\includegraphics[width=.48\columnwidth]{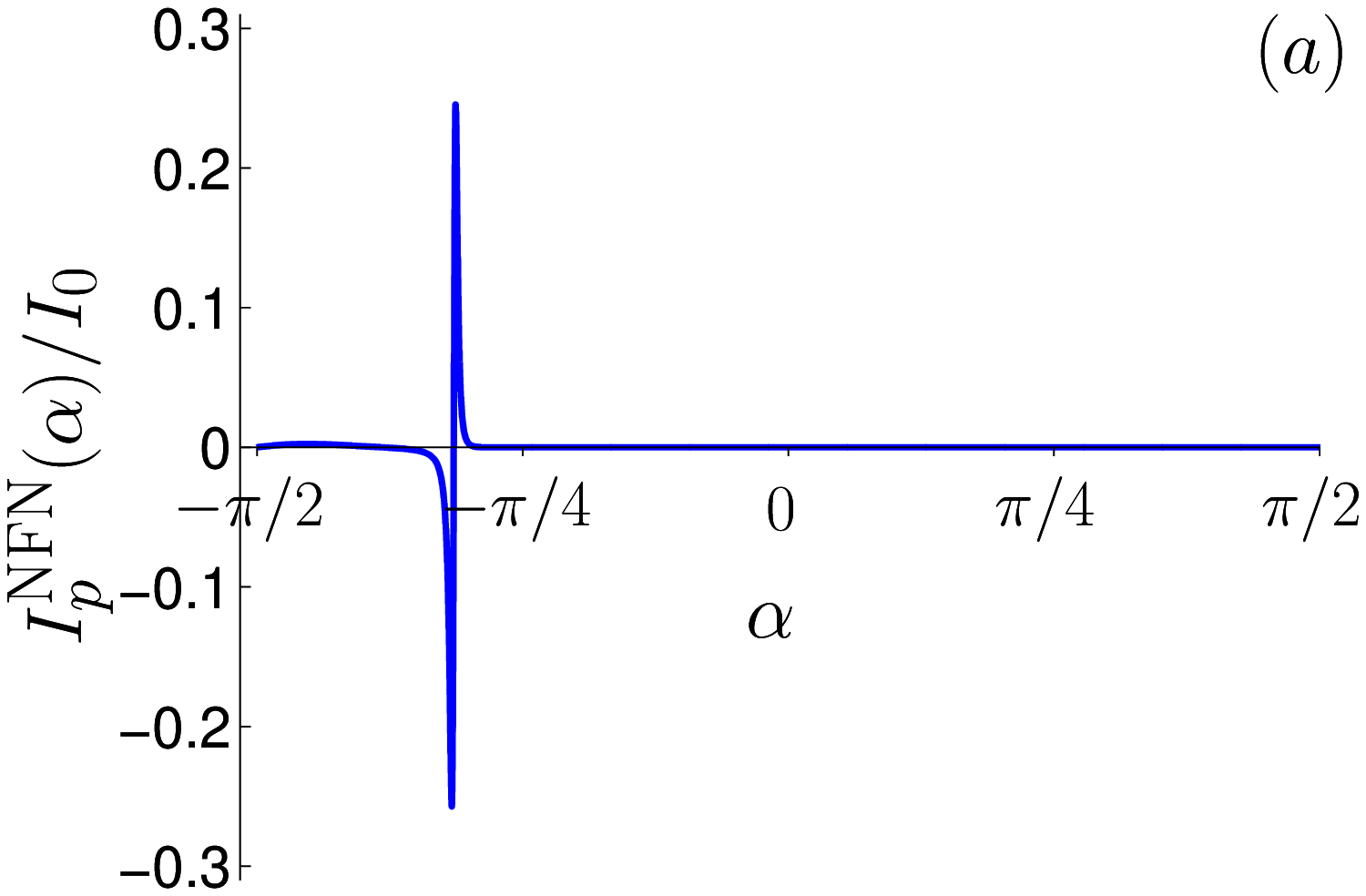}}\hfill
\subfigure{\includegraphics[width=.48\columnwidth]{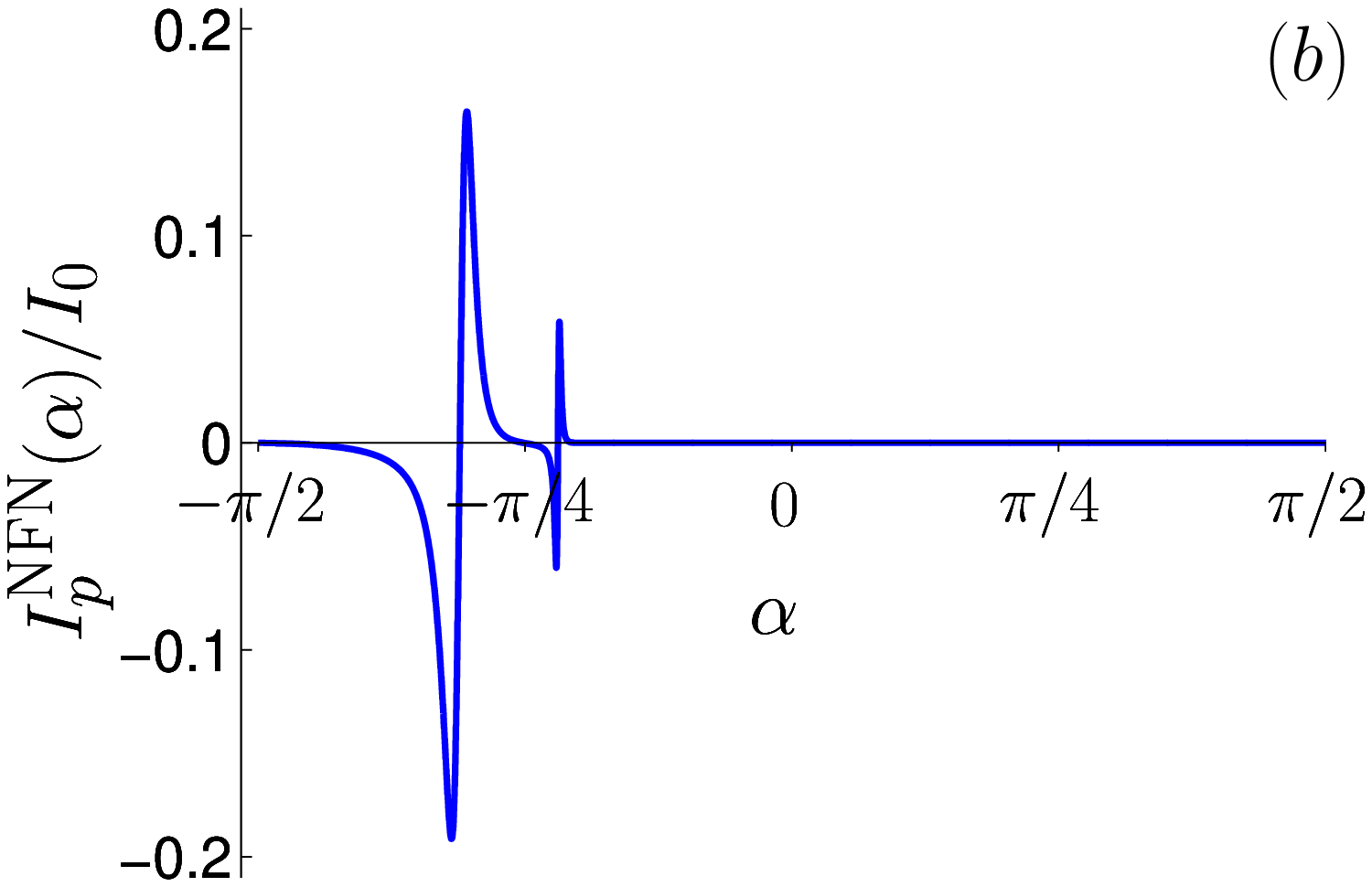}}\\ \subfigure{\includegraphics[width=.48\columnwidth]{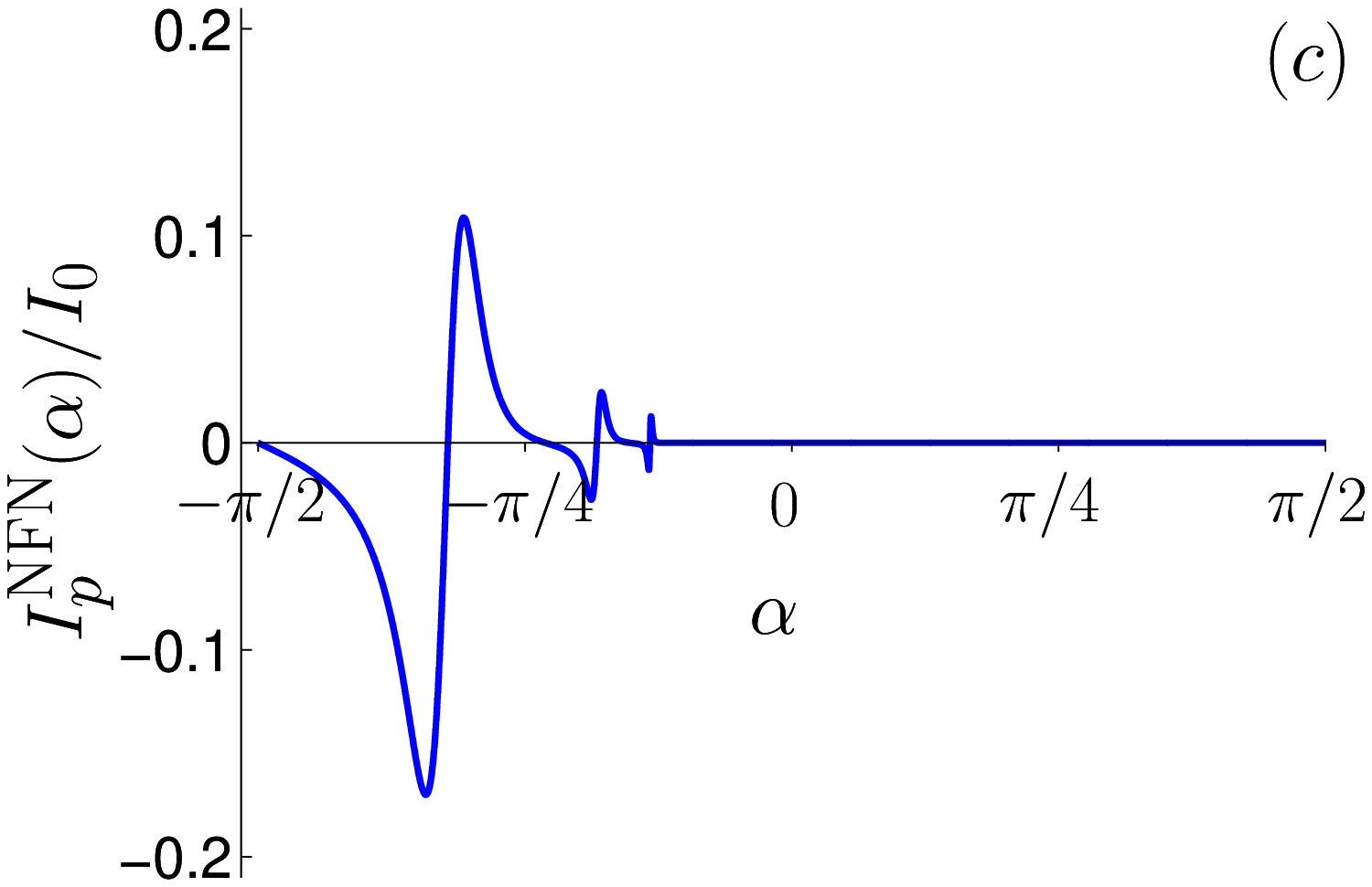}}\hfill
\subfigure{\includegraphics[width=.48\columnwidth]{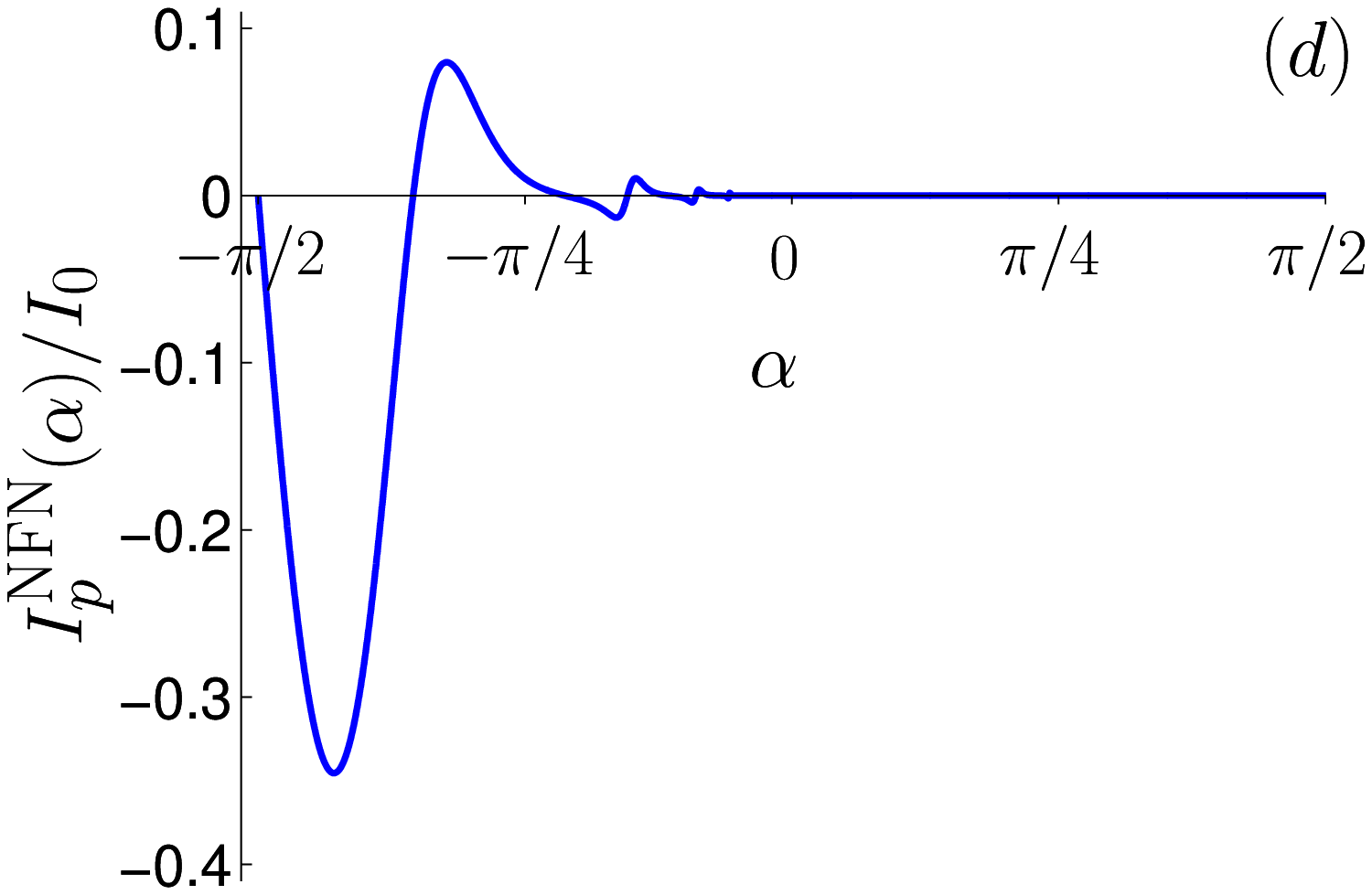}}\\ \subfigure{\includegraphics[width=.48\columnwidth]{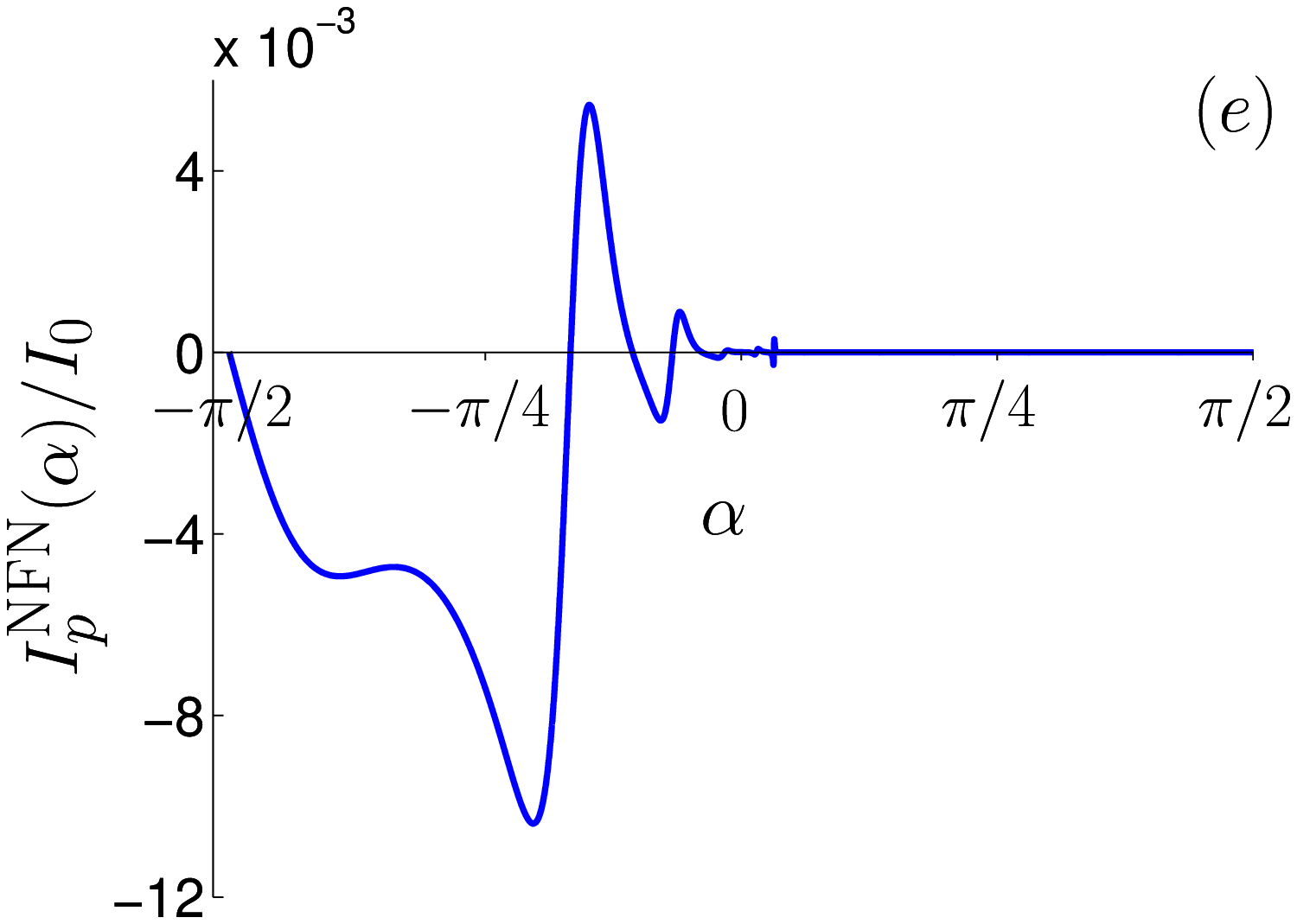}}\hfill
\subfigure{\includegraphics[width=.48\columnwidth]{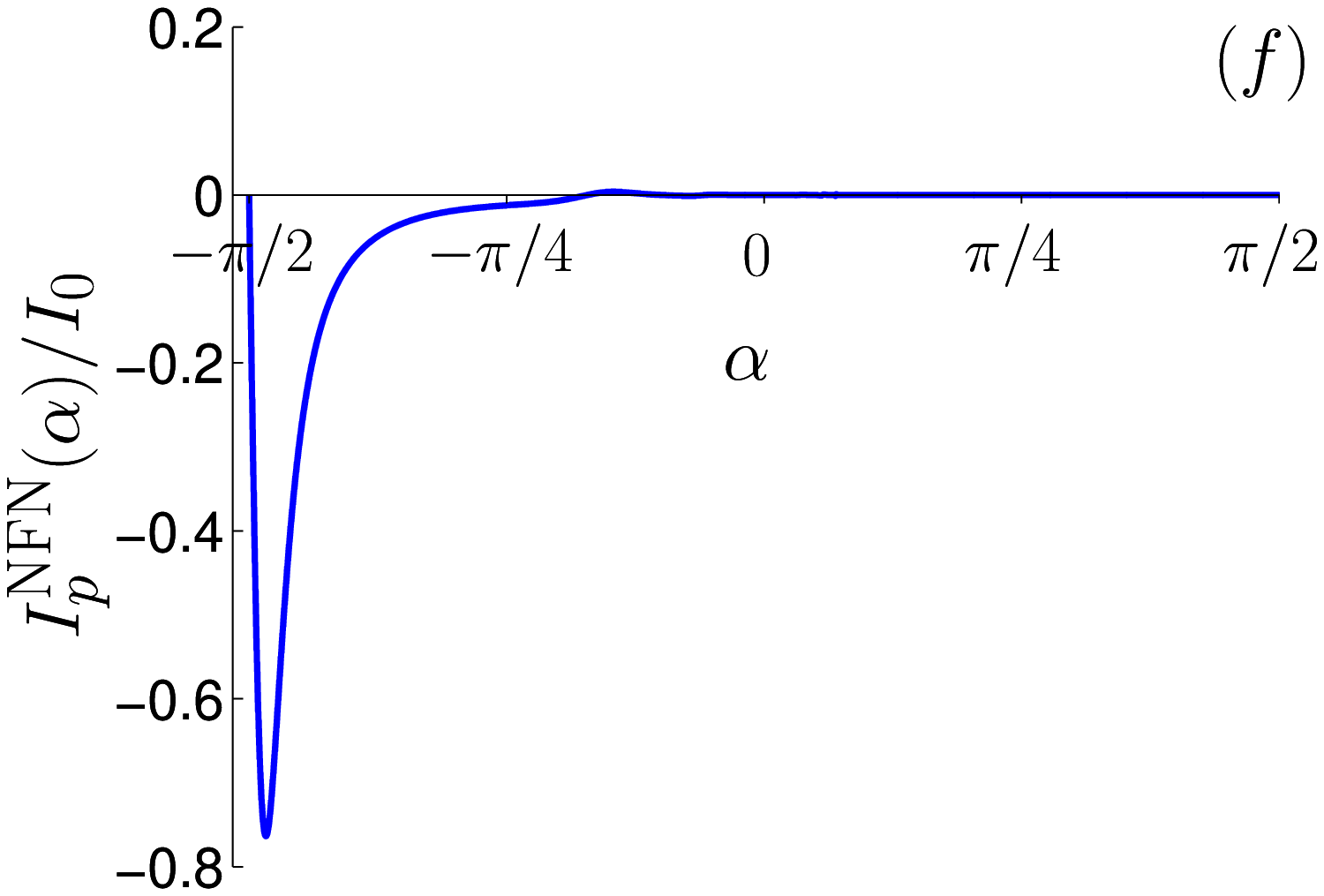}}
\caption{The pumped current $I_{p}^{\text{NFN}}(\alpha)$
  [Eq.~(\ref{eq:NFNpumpedcurrent})] as a function of the angle of
  incidence $\alpha$ for different values of energy $\epsilon/\mu$,
  (a) $\epsilon/\mu = 0.7$, (b) $\epsilon/\mu = 0.9$, (c)
  $\epsilon/\mu = 1.2$, (d) $\epsilon/\mu = 1.6$, (e) $\epsilon/\mu =
  2.4$, and (f) $\epsilon/\mu = 2.9$.  Parameters used are $\tilde{d}
  = 5$ and $\tilde{M} = 3$. }
\label{fig:IpNFNvsalpha}
\end{figure}

First we analyze the NFN pump, where the pumped current is generated
by adiabatic variation of gate voltages $V_l$ and $V_r$ which change
the chemical potential in the normal leads on the left and right of
the junction, respectively (see
Fig.~\ref{fig:NFNjunction}). Calculating the derivatives of the
reflection and transmission coefficients $r_{ll}$
[Eq.~(\ref{eq:generalreflectionLL})] and $t_{rl}$ with respect to
$\alpha_l$ and $\alpha_r$, substituting into Eq.~(\ref{eq:Pifunc}) and
using $\partial \alpha_j /(e\partial V_j) = \tan (\alpha_j)/ |\epsilon
+ \mu -eV_j|$ ($j = l, r$), the pumped current for $V_1 = V_2 \equiv
V$ and for a specific angle of incidence $\alpha$ is given by:
\begin{widetext}
\begin{equation}
  I_{p}^{\text{NFN}}(\alpha)= - I_0^{\text{NFN}}
\frac{\cos^3(\alpha_m) \sin^2 (\alpha) \cos (\alpha) \sin( 2 k_m d) } {(1 +
  \epsilon/ \mu- eV/ \mu)^2 (\cos^2 (\alpha) \cos^2(\alpha_m) \cos^2
  (k_m d) +\sin^2 (k_m d) (1 - \sin (\alpha) \sin (\alpha_m))^2)^2}.
\label{eq:NFNpumpedcurrent}
\end{equation}
\end{widetext}
 Here $I_0^{\text{NFN}} \equiv \omega e/(8\pi) \sin({\phi}) (e \delta
V_1 /\mu)(e\delta V_2/\mu)$ and 
$\sin(\alpha_m)$ is given by Eq.~(\ref{eq:alpham}).  In the limit $M
\rightarrow 0$ (i.e., $\al_m \rightarrow \al$) in an entirely normal
junction, we obtain from Eq.~(\ref{eq:NFNpumpedcurrent}) the
angle-dependent pumped current as:
\begin{equation}
  I_p^{\text{NFN}}|_{\tilde{M} =0}= I_0^{\text{NFN}}
  \int_{-\pi/2}^{\pi/2} \frac{\sin (2 k_m d) \sin^2 \alpha }{ (1 +
    \epsilon /\mu -eV /\mu ) \cos^3 \alpha} \, d\alpha.
\label{eq:NFNtotalcurrentm=0}
\end{equation}
On the other hand, the transmission $T^{\text{NFN}}(\alpha)$
[Eq.~(\ref{eq:totaltransmission})] in this limit is given by
$T^{\text{NFN}}|_{\tilde{M} \rightarrow 0} \rightarrow 1$, independent
of the angle of incidence $\alpha$. We notice that even if the
probability for transmission is one, it is possible to pump a current
in the adiabatic driving regime.  The total pumped current $I_p^{NFN}$
is then obtained by integrating over $\alpha$:
\begin{equation}
  I_p^{\text{NFN}}= \int_{-\pi/2}^{\pi/2} I_p^{\text{NFN}} (\alpha)
  \cos\alpha\, d\alpha.
\label{eq:NFNtotalcurrent}
\end{equation}
In general, this integral cannot be evaluated analytically and we have
obtained our results numerically.  
\begin{figure}
\centering
\subfigure{\includegraphics[width=.8\columnwidth]{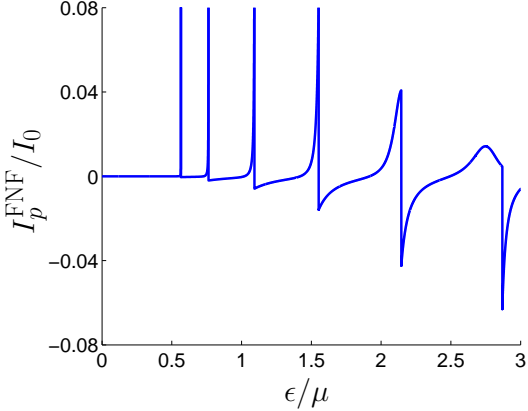}}
\\
\caption{The pumped current $I_p^{\text{NFN}}$
  [Eq.~(\ref{eq:NFNtotalcurrent})] in the NFN junction as a function
  of $\epsilon/\mu$ for $V =0$, $\tilde{d} = 5$ and $\tilde{M}= 3$. }
\label{fig:NFNIp}
\end{figure}
Figure~\ref{fig:NFNIp} shows the total pumped current
$I_p^{\text{NFN}}$ (in units of $I_0^{\text{NFN}}$) at zero bias $V =
V_1=V_2=0$ for $\tilde{M}=3$.
Comparing Figs.~\ref{fig:NFNconductance} and~\ref{fig:NFNIp} we see
that there is a correlation between the pumped current and the
conductance for the NFN junction: for low energies, the pumped current
$I_p^{NFN}$ is zero as no traveling modes are allowed in the
junction. As we increase the energy, each time a resonant mode appears
[see Eq.~(\ref{eq:resonantcondition})], the pumped current diverges
and changes sign. For energies where both positive and negative angles
of incidence contribute to the conductance, the pumped current remains
finite but keeps changing its sign. From
Figs.~\ref{fig:NFNconductance} and~\ref{fig:NFNIp} it can also be seen
that the pumped current vanishes for energies at which subsequent
resonant modes become fully transmitting. In order to gain further
insight we plot the analogue of Fig.~\ref{fig:Gvsalpha} for the pumped
current. Figure ~\ref{fig:IpNFNvsalpha} shows the pumped current
[Eq.~\ref{eq:NFNtotalcurrent}] as a function of the angle of incidence
$\alpha$ for different values of $\epsilon/\mu$. The chosen values of
$\epsilon/\mu$ are same as in Fig.~\ref{fig:Gvsalpha}. We see that the
features in Fig.~\ref{fig:IpNFNvsalpha} have a direct correlation with
the features in Fig.~\ref{fig:Gvsalpha}: whenever there is a sharp
peak in the transmission the pumped current diverges and changes
sign. The key feature that distinguishes between the pumped current
and the conductance is that the pumped current changes sign at
particular values of the energy, while the conductance does not.

Now we analyze the pumped current in the FNF junction with parallel
orientation of the magnetizations. In this system, the driving
parameters are the magnetizations $M_l$ and $M_r$ in the left and
right contacts, respectively, see Fig.~\ref{fig:FNFjunction}. After
calculating the derivatives of the reflection and transmission
coefficients, obtaining the imaginary part of Eq.~(\ref{eq:Pifunc}),
and using $\partial \alpha_j/\partial M_j = \hbar v_F /( |\epsilon +
\mu|\cos(\alpha_j) )$ ($j=l,r)$, the pumped current
$I_p^{\text{FNF}}(\alpha)$ for $M_1=M_2=M$ is:
\begin{widetext}
\begin{equation}
  I_p^{\text{FNF}}(\alpha)= - I_0^{\text{FNF}} 
\frac{\cos^3(\alpha_m) \cos(\alpha)\sin (2 k_m d)}{(1 +
  \epsilon/\mu)^2 (\cos^2 (\alpha) \cos^2 (\alpha_m) \cos^2 (k_m d)
  +\sin^2 (k_m d) (1 - \sin (\alpha) \sin (\alpha_m))^2)^2}.
\label{eq:FNFpumpedcurrent}
\end{equation}
\end{widetext}
Here $I_0^{\text{FNF}}= \omega e/(8\pi) \sin ({\phi}) (\hbar v_F\delta
M_l/\mu)(\hbar v_F \delta M_r/\mu)$ and $\sin (\alpha_m)= \sin
(\alpha) -\hbar v_F M/(|\epsilon + \mu|)$.

\begin{figure}
\subfigure{\includegraphics[width=.48\columnwidth]{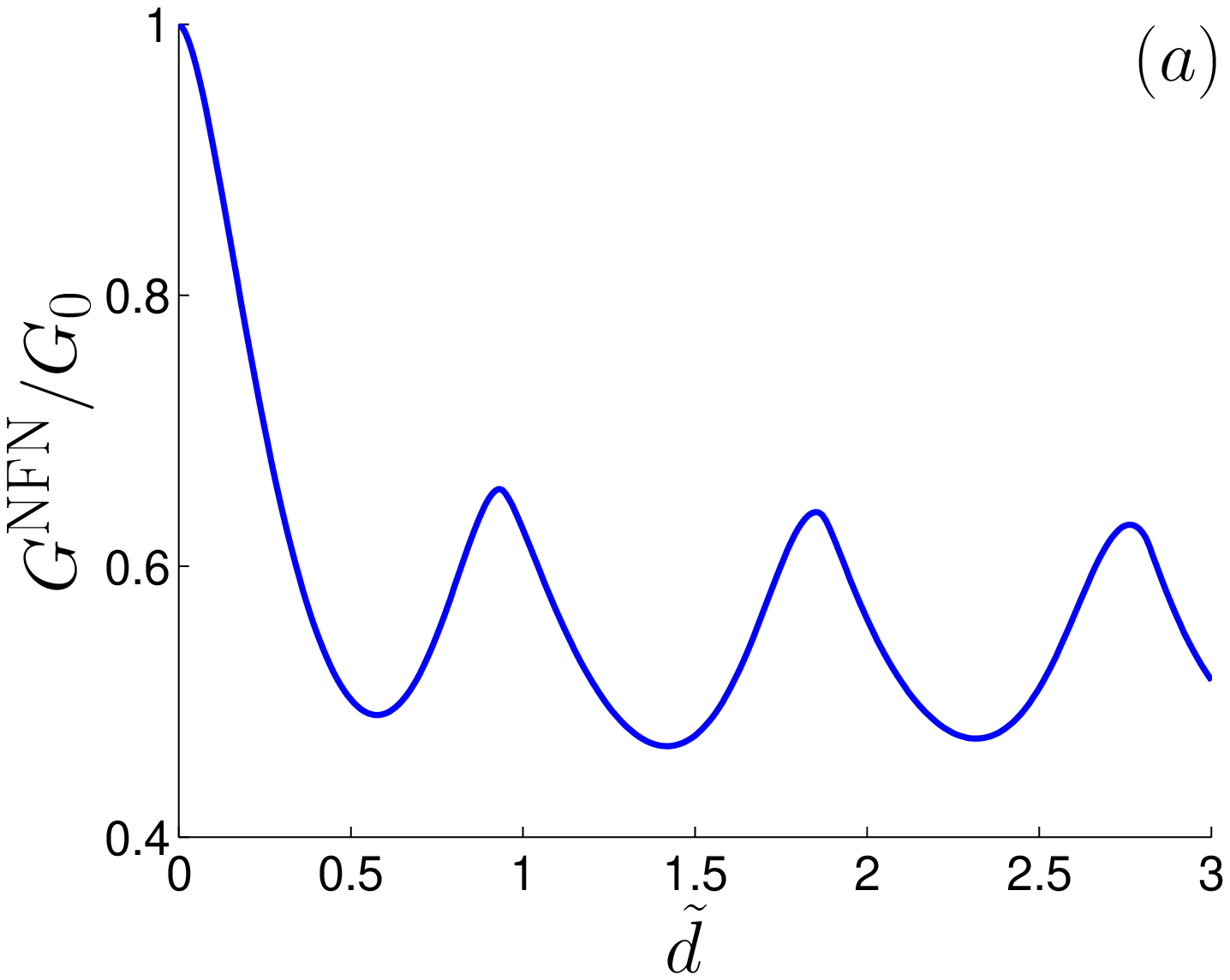}}\hfill
\subfigure{\includegraphics[width=.48\columnwidth]{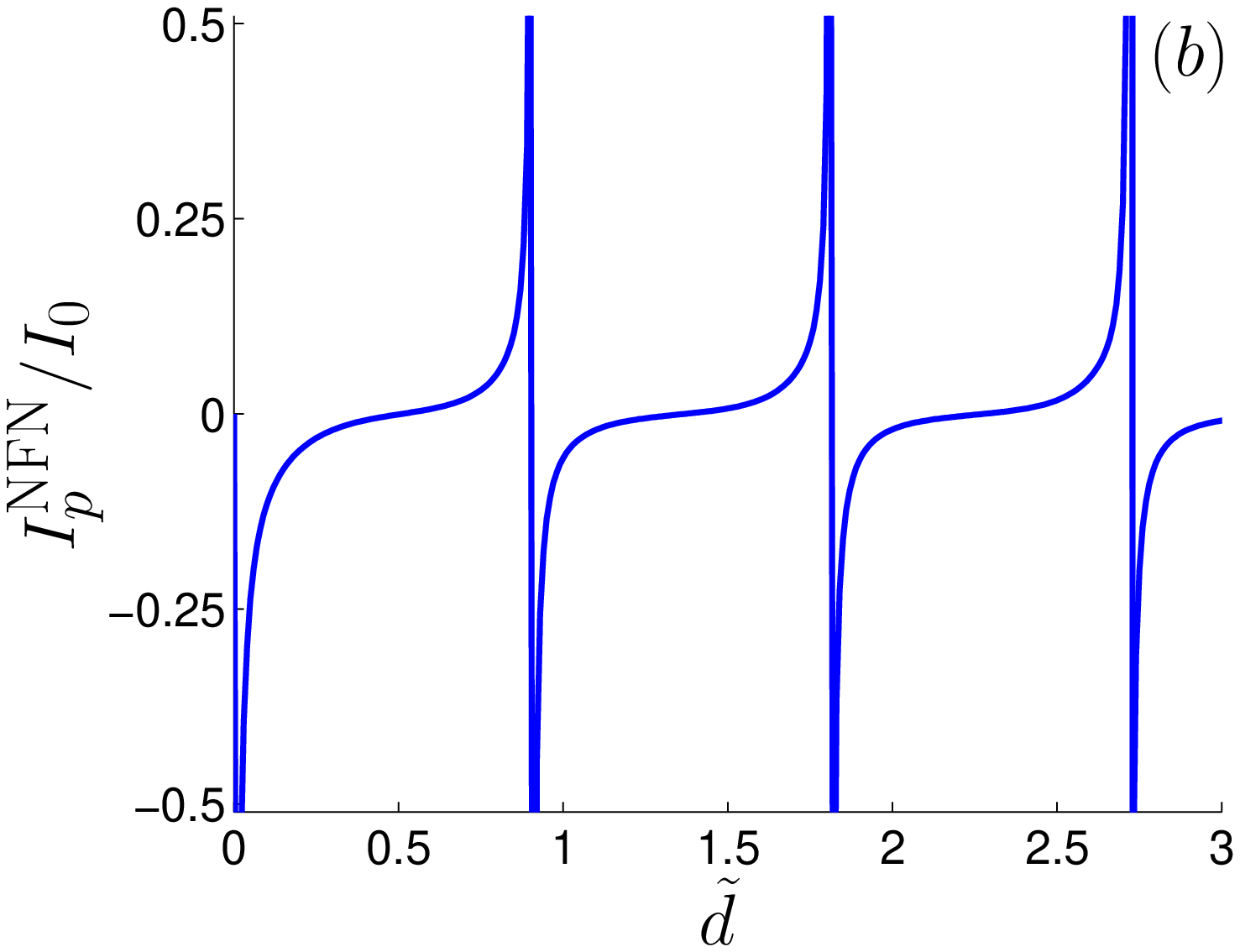}}
\caption{ (a) The conductance of the NFN junction as a function of
  $\tilde{d}$. (b) The pumped current for the NFN junction as a
  function of $\tilde{d}$. Parameters used are $\epsilon/\mu=2.5$,
  $\tilde{M} = 3$ and $V =0$.  }
\label{fig:IpNFNvsD}
\end{figure}

The behavior of the current $I_p^{\text{FNF}}$ is similar to that of
the pumped current $I_p^{\text{NFN}}$ in a NFN-junction (shown in
Fig.~\ref{fig:NFNIp}).  This can also be seen by comparing the
denominators in Eqns.~(\ref{eq:NFNpumpedcurrent}) and
(\ref{eq:FNFpumpedcurrent}). Again we observe that the pumped current
diverges at exactly the same locations where the conductance changes
sharply. But there is an important difference between both pumped
currents. The pumped current in an NFN-junction at normal incidence
vanishes, $I_p^{\text{NFN}}(\alpha=0) =0$, while
$I_p^{\text{FNF}}(\alpha=0)\ne 0$.
This difference arises because the two pumps are driven by two
different parameters (voltages in the NFN pump and magnetizations in
the FNF pump).

Finally, we briefly analyze the behavior of the pumped current as a
function of the width $d$ of the middle region. For energies below
$\epsilon_c$, the pumped current of the NFN junction decays to zero as
the width $d$ increases (there are no resonant modes in the
system). For energies larger than $\epsilon > \epsilon_c$, the pumped
current $I_p^{\text{NFN}}$ oscillates as a function of width
$\tilde{d}$. Fig.~\ref{fig:IpNFNvsD} shows the conductance and the
pumped current as a function of $\tilde{d}$ for
$\epsilon/\mu=2.5$. The peaks in the conductance correspond to the
resonance condition Eq.~(\ref{eq:resonantcondition}). The pumped
current $I_p^{\text{NFN}}$ changes its sign at exactly the same values
of $\tilde{d}$ where the conductance has a maximum.
This analysis holds as well for the FNF junction.

\section{Summary and discussion}
\label{sec:summary}

To summarize, we have analyzed quantum transport by Dirac fermion
surface states in NFN and the FNF junctions in a 3D topological
insulator. We have shown that for low energies the appearance of a new
resonant mode results in a plateau-like increment of the conductance
and a diverging pumped current in these junctions which also changes
sign.  This is our key result, and represents an experimentally
distinguishable signature between conductance and the pumped current.
We highlighted an interesting difference between the two different
pumping mechanisms for the NFN and FNF junctions, observing different
behaviors for normal incidence ($\alpha=0$). Experimentally, the NFN
pump could be realized using current technology. The FNF pump will be
more difficult to realize since it requires oscillating
magnetizations. A possible way to realize a FNF pump could be by
moving the two ferromagnetic layers coherently using a nanomechanical
oscillator~\cite{Kovalev2005}. Experimental verification of our
predictions will provide further insight into quantum transport
through these junctions.

\begin{acknowledgments}
This research was supported by the Dutch Science Foundation NWO/FOM.
\end{acknowledgments}


\begin{thebibliography}{}
\bibitem{Hasan2010} M. Z. Hasan and C. L. Kane, Rev. Mod. Phys. {\bf
  82}, 3045 (2010).
\bibitem{Konig2007} M. K\"onig, S. Wiedmann, C. Br\"une, A. Roth,
  H. Buhmann, L. W. Molenkamp, X.-L. Qi, and S.-C. Zhang, Science {\bf
    318}, 766 (2007).
\bibitem{Hsieh2008}D. Hsieh, D. Qian, L. Wray, Y. Xia, Y. S. Hor,
  R. J. Cava, and M. Z. Hasan, Nature {\bf 452}, 970 (2008) .

\bibitem{Xia2009a} Y. Xia, D. Qian, D. Hsieh, L. Wray, A. Pal, H. Lin,
  A. Bansil, D. Grauer, Y. S. Hor, R. J. Cava, and M. Z. Hasan, Nature
  Phys. {\bf 5}, 398 (2009).
\bibitem{Xia2009b} Y. Xia, D. Qian, D. Hsieh, R. Shankar, H. Lin,
  A. Bansil, A. V.  Fedorov, D. Grauer, Y. S. Hor, R. J. Cava and
  M. Z. Hasan, arXiv:0907.3089.

\bibitem{Hsieh2009a} D. Hsieh, Y. Xia, D. Qian, L. Wray, J. H. Dil,
  F. Meier, J. Osterwalder, L. Patthey, J. G. Checkelsky, N. P. Ong,
  A. V. Fedorov, H.  Lin, A. Bansil, D. Grauer, Y. S. Hor, R. J. Cava,
  and M. Z. Hasan, Nature {\bf 460}, 1101 (2009).

\bibitem{Roushan2009} P. Roushan, J. Seo, C. V. Parker, Y. S. Hor,
  D. Hsieh, D. Qian, A. Richardella, M. Z. Hasan, R. J. Cava, and
  A. Yazdani, Nature {\bf 460}, 1106 (2009).

\bibitem{Hsieh2009b} D. Hsieh, Y. Xia, L. Wray, D. Qian, A. Pal,
  J. H. Dil, J. Osterwalder, F. Meier, G. Bihlmayer, C. L. Kane,
  Y. S. Hor, R. J. Cava, and M. Z. Hasan, Science {\bf 323}, 919
  (2009).
\bibitem{HZhang2009} H. Zhang, Chao-Xing Liu, Xiao-Liang Qi, Xi
  Dai, Zhong Fang and Shou-Cheng Zhang, Nature Physics {\bf 5}, 438
  (2009).
\bibitem{Fu2008} L.\ Fu and C.\ L.\ Kane, Phys. Rev. Lett. {\bf 100},
  096407 (2008).

\bibitem{TZhang2009} T. Zhang, P. Cheng, X. Chen, J.-F. Jia, X. Ma,
  K. He, L. Wang, H. Zhang, X. Dai, Z.  Fang, X. Xie,
  Q.-K. Xue, Phys. Rev. Lett. {\bf 103}, 266803 (2009).

\bibitem{Mondal2010a} S.\ Mondal, D.\ Sen, K.\ Sengupta, and
  R.\ Shankar, Phys. Rev. Lett. {\bf 104}, 046403 (2010).

\bibitem{Mondal2010b} S. Mondal, D. Sen, K. Sengupta, and R. Shankar,
  Phys. Rev. B {\bf 82}, 045120 (2010).

\bibitem{Zhang2010} Ya Zhang and Feng Zhai, Appl. Phys. Lett. {\bf
  96}, 172109 (2010).
\bibitem{Yokoyama2010} T. Yokoyama, Y. Tanaka, and N. Nagaosa,
Phys. Rev. {\bf B} 81, 121401(R) (2010).

\bibitem{Wu2010} Zhenzua Wu, F. M. Peeters, and Kai Chang,
  Phys. Rev. B {\bf 82}, 115211 (2010).

\bibitem{Salehi2011} M. Salehi, M. Alidoust, Y. Rahnavard, and G.
  Rashedi, Physica E {\bf 43}, 4, 966 (2011).

\bibitem{Akhmerov2009} A. R. Akhmerov, J. Nilsson, and
  C. W. J. Beenakker, Phys.  Rev. Lett. {\bf 102}, 216404 (2009).

\bibitem{Tanaka2009} Y. Tanaka, T. Yokoyama, and N. Nagaosa, Phys.
  Rev. Lett. {\bf 103}, 107002 (2009).


\bibitem{Buttiker1994} M. B\"{u}ttiker, H. Thomas, and A. Pr\^{e}tre,
  Z. Phys. B {\bf 94}, 133 (1994).
\bibitem{Brouwer1998} P. W. Brouwer, Phys. Rev. B {\bf 58} 10135(R) (1998).
\bibitem{Spivak1995} B. Spivak, F. Zhou, and M. T. Beal Monod,
  Phys. Rev. B {\bf 51}, 13226 (1995).

\bibitem{Switkes1999} M. Switkes, C. M. Marcus, K. Campman, and
  A. D. Gossard, Science {\bf 283}, 1905 (1999).
\bibitem{Mucciolo2002} E. R. Mucciolo, C. Chamon, and C. M. Marcus,
  Phys. Rev. Lett. {\bf 89}, 146802 (2002).
\bibitem{Sharma2003} P. Sharma and P. W. Brouwer, Phys. Rev. Lett. {\bf
  91}, 166801 (2003).
\bibitem{Watson2003} S. K. Watson, R. M. Potok, C. M. Marcus, and
  V. Umansky, Phys. Rev. Lett. {\bf 91}, 258301 (2003).
\bibitem{Splettstoesser2005} J. Splettstoesser, Michele Governale,
  J. K\"{o}nig, and R. Fazio, Phys. Rev. Lett. {\bf 95}, 246803 (2005).
\bibitem{Sela2006} E. Sela and Y. Oreg, Phys. Rev. Lett. {\bf 96} ,
  166802 (2006).
\bibitem{Reckermann2010} F. Reckermann, J. Splettstoesser, and M. R.
  Wegewijs, Phys. Rev. Lett. {\bf 104}, 226803 (2010).

\bibitem {Prada2009} E. Prada, P. San-Jose, and H. Schomerus,
  Phys. Rev. B {\bf 80}, 245414 (2009).
\bibitem{Zhu2009} R. Zhu and H. Chen, Appl. Phys.  Lett. {\bf 95},
  122111 (2009).
\bibitem{Prada2010} E. Prada, P. San-Jose, and H. Schomerus, Solid
  State Commun. {\bf 151}, 1065 (2011).
\bibitem {Wakker2010} G.~M.~M. Wakker and\ M. Blaauboer, Phys. Rev. B,
  {\bf 82}, 205432 (2010).
\bibitem{Tiwari2010} R. P. Tiwari and M. Blaauboer,
  Appl. Phys. Lett. {\bf 97}, 243112 (2010).
\bibitem{AlosPalop2011} M. Alos-Palop and M. Blaauboer,
  Phys. Rev. B {\bf 84}, 073402 (2011). 
\bibitem{Kundu2011}
A. Kundu, S. Rao, and A. Saha, Phys. Rev. B {\bf 83}, 165451 (2011). 

\bibitem{miriam2003} M. Blaauboer, Phys. Rev. B {\bf 68}, 205316 (2003).
\bibitem{Citro2011}
R. Citro, F. Romeo, and N. Andrei, ArXiv:1109.1711.

\bibitem{Giazotto2011} F. Giazotto, P. Spathis, S. Roddaro, S. Biswas,
  F. Taddei, M. Governale and L. Sorba, Nature Physics (2011).
 
\bibitem{restriction} This restriction of the angles of incidence
  comes from the fact that the minimum and the maximum value of the
  angle $\alpha_r$ for the transmitted wavefunction is $-\pi/2$ and
  $\pi/2$ respectively.
 
  
\bibitem{difference} A fundamental difference between these pumps and
  the ones in graphene~\cite{Prada2009,Zhu2009, Prada2010, Wakker2010,
    Tiwari2010, AlosPalop2011} is the nature of the spinor in the
  Hamiltonian (\ref{eq:hamiltonian0}) which in our case represents a
  real spin due to the spin-orbit interaction, while in graphene the
  spinor represents a pseudo-spin (or the sub-lattice variable).

\bibitem{Kovalev2005} A. A. Kovalev, G. E. W. Bauer, and A. Brataas,
  Phys. Rev. Lett. {\bf 94}, 167201 (2005).







\end{thebibliography}
\end{document}